\title{Spin and quadrupole contributions to the motion of astrophysical binaries}

\author{Jan Steinhoff\footnote{Email: jan.steinhoff@ist.utl.pt,
                                URL: http://jan-steinhoff.de/physics/}\\
        Centro Multidisciplinar de Astrof\'isica (CENTRA)\\
	Instituto Superior T\'ecnico (IST)\\
	Avenida Rovisco Pais 1, 1049-001 Lisboa, Portugal}

\date{March 19th, 2014}

\documentclass[11pt,english]{article}

\usepackage{graphicx}

\usepackage[usenames]{xcolor}
\usepackage{amsmath}
\usepackage{amssymb}
\usepackage{wasysym}
\allowdisplaybreaks
\def\nl{\\ & \quad}

\DeclareMathOperator{\Order}{\mathcal{O}}

\xdefinecolor{mylinkcolor}{rgb}{0,0,0.5}
\usepackage[
	bookmarksnumbered, bookmarksopen, bookmarksopenlevel=2,
	breaklinks=true,
	colorlinks=true, filecolor=mylinkcolor, citecolor=mylinkcolor,
	linkcolor=mylinkcolor, urlcolor=mylinkcolor, menucolor=mylinkcolor,
]{hyperref}
\hypersetup{
	pdftitle={Spin and quadrupole contributions to the motion of astrophysical binaries},
	pdfauthor={Jan Steinhoff}
}
\usepackage[square,comma,numbers,sort&compress,merge]{natbib}

\begin{document}
\maketitle

\begin{abstract}
Compact objects in general relativity approximately move along geodesics of spacetime.
It is shown that the corrections to geodesic motion due to spin (dipole), quadrupole, and
higher multipoles can be modeled by an extension of the point mass action.
The quadrupole contributions are discussed in detail for astrophysical objects
like neutron stars or black holes. Implications for binaries are analyzed for a
small mass ratio situation. There quadrupole effects can encode information about
the internal structure of the compact object, e.g., in principle they allow a
distinction between black holes and neutron stars, and also different equations of state
for the latter.
Furthermore, a connection between the relativistic oscillation modes of the object
and a dynamical quadrupole evolution is established.
\end{abstract}

\tableofcontents

\section{Introduction}
The problem of motion is among of the most fundamental ones in general relativity.
As a part of the present proceedings on ``Equations of Motion
in Relativistic Gravity'' this does probably not require any explanations.
The problem is addressed using multipolar approximation schemes, the most
prominent are due to Mathisson \cite{Mathisson:1937, Mathisson:2010} and
Dixon \cite{Dixon:1979}, and another one is due to Papapetrou \cite{Papapetrou:1951}.
These particular methods have in common that equations of motion for extended bodies
are derived from the conservation of energy-momentum. In the present
contribution, the focus lies on theoretical models for compact stars and black holes based
on point-particle actions. There equations of motions follow from a
variational principle instead of conservation of energy-momentum.
These point-particle actions were probably first discussed in general relativity by
Westpfahl \cite{Westpfahl:1969:2} for the case of a pole-dipole particle
and later generalized by Bailey and Israel \cite{Bailey:Israel:1975} to
generic multipoles.

However, without further justification, it is not obvious how a point-particle
action relates to an actual extended body.
Most important is the effacing principle \cite{Damour:1983:2}, which indicates
that a nonrotating star can be represented by a point mass up to a high order
within the post-Newtonian approximation. (More details on the use of point-masses
for self-gravitating bodies within this approximation can be found
in other contributions to these proceedings, see, e.g., the contribution by G.~Sch{\"a}fer.)
This suggests that extensions
of the point-mass action can serve as models for extended bodies, even in the
self-gravitating case. A similar conclusion arises from the framework
of effective field theory applied to gravitating compact bodies
\cite{Goldberger:Rothstein:2006} (which is also covered by a different
contribution to these proceedings).
Indeed, the effective action belonging
to a compact body naturally takes on the form of a point-particle action, which puts
previous works on similar actions \cite{Westpfahl:1969:2, Bailey:Israel:1975}
into a different light. This provides enough motivation for us to further elaborate
the action approach of \cite{Bailey:Israel:1975} in Sec. \ref{action}, where it is
combined with useful aspects of more recent literature
\cite{Porto:2006, Steinhoff:Schafer:2009:2, Steinhoff:2011, DeWitt:2011,
Blanchet:Buonanno:LeTiec:2012}. An application to the post-Newtonian
approximation of self-gravitating extended bodies is omitted, because various formalisms exist
for it and the aim is to highlight aspects that are independent of (and hopefully useful for)
all of them.

It is worth mentioning that the effective field theory framework offers
a machinery which can be used, at least in principle, to \emph{compute}
the effective point-particle action from a complete microphysical description of
the extended body. In practice, however,
this procedure is not viable for realistic astrophysical objects and one must
be satisfied with a more phenomenological construction of the effective action.
This is in fact analogous to other situations in physics. For instance, it is
usually admitted that thermodynamic potentials can be derived from a microscopic
description. Yet an explicit calculation is often too complicated, or the microscopic
description is even unclear. But a phenomenological construction of thermodynamic
potentials or equations of state is usually possible.
This analogy is further elaborated in Sec.\ \ref{spinquad}.
There an adiabatic quadrupole deformation due to spin \cite{Laarakkers:Poisson:1999}
is discussed.
An application to a binary system in the extreme mass ratio case is given.
Quadrupole deformation due to an external gravitational fields is discussed
in Secs.\ \ref{Qdynamic}--\ref{symetc}, both in an adiabatic
\cite{Damour:Nagar:2009:4, Binnington:Poisson:2009, Hinderer:2007}
and a dynamical situation \cite{Chakrabarti:Delsate:Steinhoff:2013:2,
Chakrabarti:Delsate:Steinhoff:2013:1}.

A main critique against point-particles arises from the
fact that Dirac delta distributions are ill-defined sources for the
nonlinear Einstein equations. But the situation changes once one softens
the Dirac delta using regularization techniques.
It is then possible to solve the field equations iteratively within some
approximation, like the post Newtonian one.
If one regards the chosen regularization prescription as a part of the phenomenological
model, then point-particles must be
accepted as viable sources in general relativity
(at least for applications within approximation schemes).
This point is further stressed in Sec.\ \ref{Qdynamicfull}.
It is important that a weak field approximation for the point-particle
mimics the field of the actual self-gravitating extended body away from the source.
(This is precisely the criterion for the phenomenological construction
of the effective point-particle source.)
Hence, though one applies the effective source to weak field approximations, e.g.,
to compute predictions for a binary, strong-field effects from the interior of
the bodies are taken into account.

The signature of spacetime is taken to be $+2$.
Units are such that the speed of light $c$ is equal to one.
The gravitational constant is denoted by $G$.
% Lower case Latin indices from the beginning of the alphabet ($a$, $b$, \dots)
% label the individual spinning objects and then consequently take on values from one to the number of objects.
We are going to utilize three different frames, denoted by different indices.
Greek indices refer to the coordinate frame,
lower case Latin indices from the beginning of the alphabet belong to a
local Lorentz frame, and upper case Latin indices from the beginning of
the alphabet denote the so called body-fixed Lorentz frame.
% Lower case Latin indices from the middle of the alphabet ($i$, $j$, \dots) are used for the spatial part
% of the mentioned frames and are running through $i = 1, 2, 3$.
% In order to distinguish the three frames when splitting them into
% spatial and time part, we write $a = (0), (i)$ for Lorentz indices
% (or $a = (0), (1), (2), (3)$ in more detail),
% $A = [0], [i]$ for the body-fixed frame, and $\mu = 0, i$ for the coordinate frame.
% Indices appearing twice in a product are implicitly summed over its index range, except
% for label indices of the objects.
Round and square brackets are used for index symmetrization
and antisymmetrization, respectively, e.g., $A^{(\mu\nu)} \equiv \frac{1}{2} (A^{\mu\nu} + A^{\nu\mu})$.
% Partial derivatives are denoted by $\partial_{\mu}$ or by a comma as an index $~_{,\mu}$.
% Similarly, the 4-dimensional covariant derivative is written as $~_{||\mu}$ and
% the induced 3-dimensional one as $~_{;i}$.
% A 3-dimensional vector is also written in boldface, e.g., $\vct{x}$.
The convention for the Riemann tensor is
% such that
% \begin{equation}
% a_{\mu||\alpha\beta} - a_{\mu||\beta\alpha} = R_{\nu\mu\alpha\beta} a^{\nu} ,
% \end{equation}
% for an arbitrary $a_{\mu}$, or
\begin{equation}
R^{\mu}{}_{\nu\alpha\beta} = \Gamma^{\mu}{}_{\nu \beta , \alpha}
	- \Gamma^{\mu}{}_{\nu \alpha , \beta}
	+ \Gamma^{\rho}{}_{\nu \beta} \Gamma^{\mu}{}_{\rho \alpha}
	- \Gamma^{\rho}{}_{\nu \alpha} \Gamma^{\mu}{}_{\rho \beta} .
\end{equation}
% A dot $\dot{~}$ over a symbol denotes the ordinary coordinate-time derivative $\frac{d}{d t}$.
% Variations and variational derivatives are denoted by a small delta, $\delta$, and
% the 4-dimensional covariant differential by $D$. The Dirac delta distribution
% is normalized as $\int d t \, \delta(t) = 1$.
% A number in round brackets also denotes a formal post-Newtonian order in terms
% of $c^{-1}$, and should not be confused with an index in the local frame
% (the meaning should always be clear from the context).
% For a 3-dimensional antisymmetric tensor, say, $J_{ij}$, we sometimes
% also use the corresponding vector, $J_i = \tfrac{1}{2} \epsilon_{ijk} J_{jk}$.

\section{Point-particle actions\label{action}}
Action principles for spinning point particles have a long tradition, see, e.g.,
\cite{Goenner:Westpfahl:1967, Romer:Westpfahl:1969, Westpfahl:1969:2,
Hanson:Regge:1974, Bailey:Israel:1975, Leclerc:2005, Porto:2006, Natario:2008, Steinhoff:Schafer:2009:2,
Steinhoff:2011, DeWitt:2011, Blanchet:Buonanno:LeTiec:2012}.
In this section, the advantages from several of these references are brought together.
Our approach is most similar to \cite{Bailey:Israel:1975}. Compared to the
presentation in \cite{Steinhoff:2011}, a simpler (manifestly covariant)
variation technique is applied and the transition to tetrad gravity is discussed at a later
stage. This makes the derivation more transparent.

\subsection{Manifestly covariant variation}
Before we start to formulate the action principle, let us introduce a useful notation due to B.~S.~DeWitt \cite{DeWitt:1964}, see also \cite[appendix A]{DeWitt:2011}.
One can define a linear operator $G^{\nu}{}_{\mu}$ such that the covariant derivative
$\nabla_{\alpha}$ and the Lie derivative $\mathcal{L}_{\xi}$ read
\begin{equation}
\nabla_{\alpha} := \partial_{\alpha} + \Gamma^{\mu}{}_{\nu\alpha} G^{\nu}{}_{\mu} , \qquad
\mathcal{L}_{\xi} := \xi^{\mu} \partial_{\mu}
	- ( \partial_{\nu} \xi^{\mu} ) G^{\nu}{}_{\mu} .
\end{equation}
For instance, $G^{\nu}{}_{\mu}$ operates on a tensor $T_{\alpha}{}^{\beta}$ as
$G^{\nu}{}_{\mu} T_{\alpha}{}^{\beta} := - \delta^{\nu}_{\alpha} T_{\mu}{}^{\beta} + \delta^{\beta}_{\mu} T_{\alpha}{}^{\nu}$.
That is, $G^{\nu}{}_{\mu}$ is a linear operator that acts on the spacetime indices of a tensor.
Notice that $G^{\nu}{}_{\mu}$ does not act on indices of the body-fixed frame.
Further, $G^{\nu}{}_{\mu}$ obeys a product rule like
a differential operator. Similarly, we can construct a covariant differential $D$ and
a covariant variation $\Delta$ of quantities defined along a worldline $z^{\alpha}$ by
\begin{equation}
D := d + \Gamma^{\mu}{}_{\nu\alpha} (d z^{\alpha}) G^{\nu}{}_{\mu} , \qquad
\Delta := \delta + \Gamma^{\mu}{}_{\nu\alpha} (\delta z^{\alpha}) G^{\nu}{}_{\mu} .
\label{covVar}
\end{equation}
For scalars the contributions from the connection vanish. Notice that a variation of
the worldline $\delta z^{\alpha}$ is not manifestly covariant if the component values of
tensors defined on the worldline are held fixed. The variation $\Delta$ instead
parallel transports to the varied worldline. When it is applied to a tensor
\emph{field} taken at the worldline, e.g., $T_{\alpha}{}^{\beta}(z)$, then
the variation $\delta$ splits into a part due to the shift of the worldline $\delta z^{\rho}$
and a part coming from the variation of the field itself. Let us denote the latter
part by $\delta_z T_{\alpha}{}^{\beta} := (\delta T_{\alpha}{}^{\beta})(z)$,
so we have
\begin{equation}
\delta \equiv \delta_z + (\delta z^{\rho}) \partial_{\rho} , \qquad
\Delta \equiv \delta_z + (\delta z^{\alpha}) \nabla_{\alpha} , \label{covVar2}
\qquad \text{(for fields).}
\end{equation}
For instance, the metric compatibility of $\nabla_{\alpha}$ then leads to
\begin{equation}\label{Dmetric}
\Delta g_{\mu\nu} = \delta_z g_{\mu\nu} .
\end{equation}

\subsection{Action principle}
We envisage an action principle localized on a worldline $z^{\rho}(\lambda)$.
Here $\lambda$ is an arbitrary parameter, not
necessarily identical to the proper time $\tau$. (Let us require that the action is
invariant under reparametrizations of the worldline).
We further assume that the action is varied with respect to a ``body-fixed''
frame defined by Lorentz-orthonormal basis vectors $\Lambda_A{}^{\mu}(\lambda)$
labeled by $A$,
\begin{equation}\label{Lcond}
\Lambda_A{}^{\mu} \Lambda_B{}^{\nu} g_{\mu\nu} \equiv \eta_{AB} , \qquad
\Lambda_A{}^{\mu} \Lambda_B{}^{\nu} \eta^{AB} \equiv g^{\mu\nu} .
\end{equation}
Now, stars are in general differentially rotating and it is difficult to interpret a body-fixed frame.
Such a frame is thus rather an abstract element of our theoretical model,
inspired by the Newtonian theory of rigid bodies (see, e.g., \cite[Sec.\ 3.1.1]{Steinhoff:2011}).

The constraint (\ref{Lcond}) implies that $\Lambda_A{}^{\mu}$ and $g_{\mu\nu}$
are in general not independent and one
should take special care when both are varied at the same time. In order to address
this problem, we split the variation $\Delta \Lambda_A{}^{\nu}$ as
\begin{align}
\Lambda^{A\mu} \Delta \Lambda_A{}^{\nu} = \Lambda^{A[\mu} \Delta \Lambda_A{}^{\nu]}
	+ \frac{1}{2} \Delta ( \Lambda^{A\mu} \Lambda_A{}^{\nu} )
= \Delta \Theta^{\mu\nu} - \frac{1}{2} g^{\mu\alpha} g^{\nu\beta} \delta_z g_{\alpha\beta} . \label{varLambdasplit}
\end{align}
where we used $\delta g^{\mu\nu} = - g^{\mu\alpha} g^{\nu\beta} \delta g_{\alpha\beta}$
and (\ref{Dmetric}).
In the last step, we also introduced the abbreviation
\begin{equation}
\Delta \Theta^{\mu\nu} := \Lambda^{A[\mu} \Delta \Lambda_A{}^{\nu]} ,
\end{equation}
which is similar to the antisymmetric variation symbol used in
\cite{Hanson:Regge:1974}, see also \cite[Eq.\ (2.7)]{Blanchet:Buonanno:LeTiec:2012}.
The independence of $\Delta \Theta^{\mu\nu}$ from the metric variation
$\delta_z g_{\alpha\beta}$ will be made more manifest in Sec.\ \ref{tetrad}. For now
let us just appeal to the fact that the 6 degrees of freedom of the antisymmetric symbol
$\Delta \Theta^{\mu\nu}$ exactly matches the degrees of freedom of a
Lorentz frame (3 boosts and 3 rotations). Thus $\Delta \Theta^{\mu\nu}$
corresponds to the independent variation of the body-fixed Lorentz frame.

Let us consider an action that is as generic as possible,
\begin{gather}
W = W_F + W_M , \qquad
W_F[g_{\mu\nu}, \dots] = \frac{1}{16\pi G} \int d^4 x \, \sqrt{-g} R + \dots , \\
	W_M[g_{\mu\nu}, z^{\rho}, \Lambda_A{}^{\mu}, \dots] = \int d \lambda \,
		L_M(g_{\mu\nu}, u^{\mu}, \Lambda_A{}^{\mu}, \Omega^{\mu\nu}, \phi_I) ,
\end{gather}
Here $\phi_I$ collectively denotes other dependencies of the Lagrangian $L_M$
and the dots denote other fields, like the electromagnetic one.
(In this section $I$ is a multi-index that may comprise any sort of spacetime, Lorentz, or label indices.)
Notice that fields like $g_{\mu\nu}$ are taken at the worldline position $z^{\rho}$ in $W_M$.
The 4-velocity $u^{\mu}$ and the angular velocity $\Omega^{\mu\nu}$ are defined by
\begin{align}
u^{\mu} := \frac{d z^{\mu}}{d \lambda} , \qquad
\Omega^{\mu\nu} := \Lambda^{A\mu} \frac{D \Lambda_A{}^{\nu}}{d \lambda} , \label{velos}
% 	= {\Lambda_{A}}^{a} \left[ \frac{d \Lambda^{Ab}}{d \lambda}
% 		+ \Lambda^{Ac} \Gamma^b{}_{cd}(z^e) u^d \right] , \\
% 	&= {\Lambda_{A}}^{a} \frac{d \Lambda^{Ab}}{d \tau}
% 		+ \Gamma^{ba}{}_{c}(z^e) u^c ,
\end{align}
Notice that $\Omega^{\mu\nu}$ is antisymmetric due to (\ref{Lcond}) and
$D g_{\mu\nu} / d \lambda = 0$.

\subsection{Variation}
For the sake of deriving equations of motion, we may assume $\delta \lambda = 0$.
Then the variation can be commuted with ordinary or partial $\lambda$-derivatives.
Furthermore, the Lagrangian $L_M$ is a scalar and we can make use of
$\delta L_M \equiv \Delta L_M$ to write its variation in a manifestly covariant manner,
\begin{equation}\label{LMvar2}
\delta L_M = p_{\mu} \Delta u^{\mu}
	+ \frac{1}{2} S_{\mu\nu} \Delta \Omega^{\mu\nu}
	+ \frac{\partial L_M}{\partial \Lambda_A{}^{\mu}} \Delta \Lambda_A{}^{\mu}
	+ \frac{\partial L_M}{\partial g_{\mu\nu}} \Delta g_{\mu\nu}
	+ \frac{\partial L_M}{\partial \phi_I} \Delta \phi_I ,
\end{equation}
where we have defined the linear momentum $p_{\mu}$ and spin $S_{\mu\nu}=-S_{\nu\mu}$ as
generalized momenta belonging to the velocities $u^{\mu}$ and $\Omega^{\mu\nu}$,
\begin{equation}\label{Lder}
p_\mu := \frac{\partial L_M}{\partial u^{\mu}} , \qquad
S_{\mu\nu} := 2 \frac{\partial L_M}{\partial \Omega^{\mu\nu}} .
\end{equation}
It should be noted that (\ref{LMvar2}) can be checked using a usual
variation $\delta$ together with the identity (\ref{LMidentity}), but
here it is a simple consequence of the chain rule for $\Delta$.
Obviously this method nicely organizes the Christoffel symbols.

The 5 individual terms in (\ref{LMvar2}) are transformed as follows:
\begin{itemize}
\item The 1st term of (\ref{LMvar2}) is evaluated with the help of
\begin{equation}
\Delta u^{\mu} \equiv \delta u^{\mu} + \Gamma^{\mu}{}_{\alpha\beta} u^{\alpha} \delta z^{\beta}
        = \frac{D \delta z^{\mu}}{d \lambda} . \label{uvar}
\end{equation}
\item The 2nd term of (\ref{LMvar2}) requires the most work. In order to
evaluate $\Delta \Omega^{\mu\nu}$, we need to commute $\Delta$ with the
covariant differential $D$ contained in $\Omega^{\mu\nu}$, Eq.~(\ref{velos}).
The definitions in (\ref{covVar}) lead to
\begin{equation}
[ \Delta, D ]
% = ( \delta \Gamma^{\mu}{}_{\nu\alpha} ) ( d z^{\alpha} ) G^{\nu}{}_{\mu}
%         - ( d \Gamma^{\mu}{}_{\nu\beta} ) (\delta z^{\beta}) G^{\nu}{}_{\mu}
%         + \Gamma^{\mu}{}_{\nu\beta} \Gamma^{\rho}{}_{\sigma\alpha} ( d z^{\alpha} ) ( \delta z^{\beta} )
%                 [ G^{\nu}{}_{\mu}, G^{\sigma}{}_{\rho} ]
= [ (\delta_z \Gamma^{\mu}{}_{\nu\alpha}) - (\delta z^\beta) R^{\mu}{}_{\nu\alpha\beta} ]
        ( d z^{\alpha} ) G^{\nu}{}_{\mu} . \label{commuteDD}
\end{equation}
Notice the analogy to the commutator of covariant derivatives, which also
gives rise to curvature.
It is useful to derive intermediate commutators first, for instance
\begin{equation}
[ G^{\nu}{}_{\mu} , G^{\beta}{}_{\alpha} ]
        = \delta_{\mu}^{\beta} G^{\nu}{}_{\alpha} - \delta_{\alpha}^{\nu} G^{\beta}{}_{\mu} .
\end{equation}
Next, we express $\delta_z \Gamma^{\mu}{}_{\nu\alpha}$ in (\ref{commuteDD})
with the help of
\begin{equation}
\delta \Gamma^{\nu}{}_{\beta\alpha} = \frac{1}{2} g^{\nu\rho} \left[
	 \nabla_{\beta} \delta g_{\alpha\rho} + \nabla_{\alpha} \delta g_{\beta\rho}
	 - \nabla_{\rho} \delta g_{\alpha\beta} \right] . \label{Gvar}
\end{equation}
Now it is straightforward to evaluate $\Delta \Omega^{\mu\nu}$.
In the result, we replace $\Delta \Lambda_A{}^{\mu}$ using (\ref{varLambdasplit}),
make use of
\begin{equation}
\frac{D \delta_z g_{\rho\sigma}}{d \lambda}
= u^{\alpha} (\nabla_{\alpha} \delta g_{\rho\sigma})(z) ,
% = u^{\alpha} g_{\beta(\rho} \delta_z \Gamma^{\beta}{}_{\sigma)\alpha} .
\end{equation}
% It is most convenient to check this relation using (\ref{Gvar}).
and finally arrive at
\begin{equation}
\begin{split}
\Delta \Omega^{\mu\nu}
% &= \Delta \Lambda_A{}^{\mu} \frac{D \Lambda^{A\nu}}{d \lambda}
% 	+ \Lambda_A{}^{\mu} \Delta \frac{D \Lambda^{A\nu}}{d \lambda} \\
&= \frac{D( \Delta \Theta^{\mu\nu} )}{d \lambda}
	+ 2 \Omega_{\alpha}{}^{[\mu} \Delta \Theta^{\nu]\alpha}
	+ R^{\mu\nu}{}_{\alpha\beta} u^{\alpha} \delta z^{\beta} \nl
	+ \Omega^{\alpha[\mu} g^{\nu]\beta} \delta_z g_{\alpha\beta}
% 	+ u^{\alpha} g^{\beta[\mu} \delta_z \Gamma^{\nu]}{}_{\beta\alpha}
	+ g^{\beta[\mu} g^{\nu]\rho} u^{\alpha} (\nabla_{\beta} \delta g_{\rho\alpha})(z) . \label{Ovar}
\end{split}
\end{equation}
\item Before proceeding to the 3rd term of (\ref{LMvar2}), let us recall
the transformation property of a tensor under an infinitesimal
coordinate transformation
$x^{\mu'} = x^{\mu} - \xi^{\mu}$,
\begin{equation}
\phi_{I'} - \phi_I = - ( \partial_{\nu} \xi^{\mu} ) G^{\nu}{}_{\mu} \phi_I , \quad
\text{e.g., } \quad
u^{\mu'} - u^{\mu} = - u^{\nu} \partial_{\nu} \xi^{\mu} .
\end{equation}
The Lagrangian is a scalar and thus invariant, but it depends on tensors which
transform. As $\partial_{\nu} \xi^{\mu}$ is quite arbitrary, the invariance of the Lagrangian $L_M$ leads to the identity
\begin{equation}\label{LMidentity}
p_{\mu} u^{\nu} + S_{\mu\alpha} \Omega^{\nu\alpha}
+ \frac{\partial L_M}{\partial \Lambda_A{}^{\mu}} \Lambda_A{}^{\nu}
- 2 \frac{\partial L_M}{\partial g_{\nu\alpha}} g_{\mu\alpha}
+ \frac{\partial L_M}{\partial \phi_I} G^{\nu}{}_{\mu} \phi_I \equiv 0 .
\end{equation}
We eliminate the partial derivative of $L_M$ with respect to $\Lambda_A{}^{\mu}$
using this relation and we replace $\Delta \Lambda_A{}^{\mu}$ using
(\ref{varLambdasplit}) to arrive at
\begin{align}
\frac{\partial L_M}{\partial \Lambda^{A\mu}} \Delta \Lambda^{A\mu}
% = \frac{\partial L_M}{\partial \Lambda^{B\nu}} \Lambda^{B}{}_{\mu} \Lambda^{A\mu} \Delta \Lambda_A{}^{\nu}
% = \frac{\partial L_M}{\partial \Lambda^{B\nu}} \Lambda^{B}{}_{\mu} \left[ \Delta \Theta^{\mu\nu} - \frac{1}{2} g^{\mu\alpha} g^{\nu\beta} \delta_z g_{\alpha\beta} \right]
&= \frac{1}{2} \left[ p^{\mu} u^{\nu} - S_{\alpha}{}^{\mu} \Omega^{\nu\alpha}
        + (G^{\mu\nu} \phi_I) \frac{\partial L_M}{\partial \phi_I}
        - 2 \frac{\partial L_M}{\partial g_{\mu\nu}}
        \right] \delta_z g_{\mu\nu} \nonumber \nl
+ \left[ p_{\mu} u_{\nu} - S_{\alpha\mu} \Omega_{\nu}{}^{\alpha}
        - (G_{\mu\nu} \phi_I) \frac{\partial L_M}{\partial \phi_I} \right]
        \Delta \Theta^{\mu\nu} .
\end{align}
\item In the 4th term of (\ref{LMvar2}) we use (\ref{Dmetric}).
\item The 5th term of (\ref{LMvar2}) is not touched for now, as this requires
a specialization of $\phi_I$. This is discussed in the Sec.\ \ref{EOMassumptions}.
\end{itemize}
All these transformations are now applied to (\ref{LMvar2}).
Furthermore, we insert a unity in the form of
\begin{equation}
1 \equiv \int d^4 x \, \delta_{(4)}, \qquad
\delta_{(4)} := \delta(x^{\mu} - z^{\mu}) ,
\end{equation}
into the terms containing field variations of type $\delta_z$. This allows
one to rewrite these variations at the spacetime point $x^{\mu}$ and
perform partial integrations.
Notice that $\delta_{(4)}$ has compact support for finite $\lambda$-intervals, so
these partial integrations do not require assumptions on field variations at the
spatial boundary.
Finally, (\ref{LMvar2}) turns into
\begin{align}
\delta L_M &=
	\int d^4 x \bigg[
		p^{\mu} u^{\nu} \delta_{(4)}
		        + (G^{\mu\nu} \phi_I) \frac{\partial L_M}{\partial \phi_I} \delta_{(4)}
% 		        G^{\mu\nu} + S_{\alpha}{}^{\mu} \Omega^{\nu\alpha}
% 			+ \frac{1}{3} J^{\alpha\beta\rho\mu} R_{\alpha\beta\rho}{}^{\nu} \delta_{(4)}
		- \nabla_{\alpha} ( S^{\alpha\mu} u^{\nu} \delta_{(4)} )
% 		- \frac{2}{3} \nabla_{\beta} \nabla_{\alpha} ( J^{\beta\mu\nu\alpha} \delta_{(4)} )
	\bigg] \frac{\delta g_{\mu\nu}(x)}{2} \nonumber \nl
	+ \frac{\partial L_M}{\partial \phi_I} \Delta \phi_I
        + \bigg[
		p_{\mu} u_{\nu}
		- (G_{\mu\nu} \phi_I) \frac{\partial L_M}{\partial \phi_I}
		- \frac{1}{2} \frac{D S_{\mu\nu}}{d \lambda}
	\bigg] \Delta \Theta^{\mu\nu} \label{LMvar} \nl
	+ \bigg[
		\frac{1}{2} S_{\alpha\beta} R^{\alpha\beta}{}_{\rho\mu} u^{\rho}
% 		+ \frac{1}{6} J^{\rho\nu\alpha\beta} \nabla_{\mu} R_{\rho\nu\alpha\beta}
                - \frac{D p_\mu}{d \lambda}
	\bigg] \delta z^{\mu}
	+ \frac{d}{d \lambda} \bigg[ p_{\mu} \delta z^{\mu}
		+ \frac{1}{2} S_{\mu\nu} \Delta \Theta^{\mu\nu}
	\bigg] . \nonumber
\end{align}
Notice that the covariant and ordinary derivatives
with respect to $\lambda$ are identical for the last term.

\subsection{Metric versus tetrad gravity\label{tetrad}}
The separation of metric and body-fixed-frame variations by means of
(\ref{varLambdasplit}) is an elegant trick to derive equations of motion.
However, if further calculations at the level of the action are performed,
one often needs an explicit split between gravitational and body-fixed-frame degrees
of freedom. This can be achieved by introducing a tetrad gravitational field
$e_a{}^{\mu}(x)$, that is, a field of Lorentz-orthonormal basis vectors labeled by $a$
and defined at every spacetime point $x^{\rho}$,
\begin{equation}
e_a{}^{\mu} e_b{}^{\nu} g_{\mu\nu} \equiv \eta_{ab} , \qquad
e_a{}^{\mu} e_b{}^{\nu} \eta^{ab} \equiv g^{\mu\nu} .
\end{equation}
Tetrad gravity replaces the metric by virtue of the latter relation and regards
$e_a{}^{\mu}$ as the fundamental gravitational field in the variation principle.
This allows us to split $\Lambda_A{}^{\mu}$ as
\begin{equation}\label{Lsplit}
\Lambda_A{}^{\mu} = \Lambda_A{}^a e_a{}^{\mu}(z) ,
\end{equation}
where $\Lambda_A{}^a$ is now just a usual (flat-spacetime) Lorentz matrix
\begin{equation}
\Lambda_A{}^a \Lambda_B{}^b \eta_{ab} \equiv \eta_{AB} , \qquad
\Lambda_A{}^a \Lambda_B{}^b \eta^{AB} \equiv \eta^{ab} .
\end{equation}
Thus $\Lambda_A{}^a$ is independent of the gravitational field $e_a{}^{\mu}$
and the announced manifest split is indeed given by (\ref{Lsplit}).

Based on this split, we can understand the meaning of $\Delta \Theta^{\mu\nu}$
in more detail.
As $\Lambda_A{}^a$ is a usual Lorentz matrix, we can follow Ref.\ \cite{Hanson:Regge:1974}
and describe its independent variations by an antisymmetric symbol
$\delta \theta^{ab} := \Lambda^{Aa} \delta \Lambda_A{}^b$.
Then $\Delta \Theta^{\mu\nu}$ reads explicitly
\begin{equation}\label{DTexplicit}
\Delta \Theta^{\mu\nu} = e_a{}^{\mu} e_b{}^{\mu} \delta \theta^{ab}
        + ( \Gamma^{[\nu\mu]}{}_{\alpha} + e^{a[\mu} \partial_{\alpha} e_a{}^{\nu]} ) \delta z^{\alpha}
        + e^{a[\mu} \delta_z e_a{}^{\nu]} .
\end{equation}
One can be even more explicit and write $\delta \theta^{ab}$ as a linear combination of
six independent variations of angle variables parameterizing $\Lambda^{Aa}$,
see \cite[Sec.\ 3.A]{Hanson:Regge:1974}.
Anyway, $\Delta \Theta^{\mu\nu}$ is in fact a linear combination of
the independent frame variations $\delta \theta^{ab}$ with other variations.
Now, it is legitimate to regard $\Delta \Theta^{\mu\nu}$, $\delta z^{\alpha}$, and $\delta e_a{}^{\mu}$
as independent variations instead of $\delta \theta^{ab}$, $\delta z^{\alpha}$, and $\delta e_a{}^{\mu}$.
This just corresponds to a linear recombining of the equations of motion following from the variation.
Equation (\ref{DTexplicit}) shows that this recombination manifestly removes noncovariant terms
related to the $\delta z^{\alpha}$-variation and an antisymmetric part of the energy-momentum tensor
due to $e^{a[\mu} \delta_z e_a{}^{\nu]}$ (the symmetric part arises from
$e^{a(\mu} \delta e_a{}^{\nu)} = \frac{1}{2} \delta g^{\mu\nu}$ as usual).
All of this is important for the next section, where equations of motion
are deduced from (\ref{LMvar}).

In the next step it is possible to return to metric gravity by a partial
gauge fixing of the tetrad.
For instance, a possible gauge condition is to require that the matrix $(e_{a\mu})$
is symmetric (in spite of the different nature of its indices). Then
$e_a{}^{\mu}$ is given by the matrix square-root of the metric.
This gauge choice leads to the same conclusions as in \cite[Sec.\ IV.B]{Porto:2006},
where a more direct construction was followed.
In the end, the partially gauge-fixed tetrad is a function of the metric, so we have obtained a
metric gravity theory.
It might look like the introduction of a tetrad field accompanied by an enlarged gauge group
of gravity is just extra baggage. However, more gauge freedom is important for applications,
as gauges can and should be adopted to the problem at hand. For instance, for an
ADM-like canonical formulation of spinning particles, it is a wise choice to adopt the Schwinger
time-gauge for the tetrad \cite{Steinhoff:Schafer:2009:2}. Further, some subtle
aspects of the consistency of the theory can be analyzed more easily within tetrad
gravity (e.g., the algebra of gravitational constraints, because after reduction to metric gravity
the gravitational field momentum receives complicated corrections \cite{Steinhoff:Schafer:2009:2}).
Spinning particles should always be coupled to tetrad gravity in the first place.

\section{Equations of motion}
In order to draw conclusions from (\ref{LMvar}), one must further specialize
the so far arbitrary $\phi_I$. The assumptions we are going to introduce in
the following are not the least restrictive, but already allow important
insights on the structure of the equations of motion.

\subsection{Further assumptions\label{EOMassumptions}}
Let us assume from now on that the $\phi_I$ can be split into two groups.
We denote by $\phi_I^{\text{field}}$ the part that contains spacetime fields
(functions of $x$), so its variation $\Delta \phi_I^{\text{field}}$ can be
evaluated using (\ref{covVar2}). The second group $\phi_I^{\text{wl}}$
contains variables defined on the worldline only (functions of $\lambda$)
and its first order derivatives $\dot{\phi}_I^{\text{wl}}$, where $\dot{~} := D / d \lambda$.
Most importantly, we assume that the $\delta \phi_I^{\text{wl}}$ correspond to independent variational
degrees of freedom, like Lagrange multipliers or the dynamical multipoles
introduced in Sec.\ \ref{Qdynamic}. Without loss of generality, one can then
assume that the $\phi_I^{\text{wl}}$ carry indices of the body-fixed frame
instead of spacetime indices, so that $G^{\nu}{}_{\mu} \phi_I^{\text{wl}} = 0$.

Notice that our assumptions do not allow time (i.e., $\lambda$) derivatives of
$u^{\mu}$ and $\Omega^{\mu\nu}$ as part of the $\phi_I^{\text{wl}}$.
If such accelerations would appear in subleading contributions of the Lagrangian
(within some approximation scheme), then one can often remove them by a redefinition
of variables \cite{Damour:Schafer:1991}. Further, acceleration-dependent
Lagrangians are often problematic due to Ostrogradsky instability.
For these reasons, we also assumed that
at most first-order time derivatives of the $\phi_I^{\text{wl}}$ appear in $L_M$.
However, our assumptions here are not entirely exhaustive. For instance,
a concrete situation for which our assumptions should be relaxed in the future is
discussed at the end of Sec.\ \ref{construct}.

\subsection{Equations of motion for linear momentum and spin}
With these assumptions, we have
\begin{equation}
\begin{split}
\frac{\partial L_M}{\partial \phi_I} \Delta \phi_I
&= \frac{\partial L_M}{\partial \phi_I^{\text{field}}}
        [ \delta_z \phi_I^{\text{field}}
        + (\delta z^{\alpha}) \nabla_{\alpha} \phi_I^{\text{field}} ] \nl
+ \left[ \frac{\partial L_M}{\partial \phi_I^{\text{wl}}}
        - \frac{d \psi^I_{\text{wl}}}{d \lambda} \right] \delta \phi_I^{\text{wl}}
+ \frac{d}{d \lambda} \left[ \psi^I_{\text{wl}} \delta \phi_I^{\text{wl}} \right] , \label{phidecompose}
\end{split}
\end{equation}
where we used that the worldline variables do not carry spacetime indices
and we introduced their canonical generalized momenta,
\begin{equation}
\psi^I_{\text{wl}} := \frac{\partial L_M}{\partial \dot{\phi}_I^{\text{wl}}} .
\end{equation}
The second line leads to the usual Euler-Lagrange equtions for the
worldline degrees of freedom $\phi_I^{\text{wl}}$, which are discussed in
Sec.\ \ref{Qdynamicfull}. Let us focus on the other terms for now. Using the
arbitrariness and independence of $\delta z^{\mu}$ and $\Delta \Theta^{\mu\nu}$,
we can read off the equations of motion for the linear momentum
and the spin from (\ref{LMvar}) and (\ref{phidecompose}),
\begin{align}
\frac{D p_\mu}{d \lambda} &= \frac{1}{2} S_{\alpha\beta} R^{\alpha\beta}{}_{\rho\mu} u^{\rho} + (\nabla_{\mu} \phi_I^{\text{field}}) \frac{\partial L_M}{\partial \phi_I^{\text{field}}} , \label{pEOM} \\
\frac{D S_{\mu\nu}}{d \lambda} &= 2 p_{[\mu} u_{\nu]}
		- 2 (G_{[\mu\nu]} \phi_I^{\text{field}}) \frac{\partial L_M}{\partial \phi_I^{\text{field}}} . \label{SEOM}
\end{align}
The total $\lambda$-derivative in the last line of (\ref{LMvar}) was ignored here.
(Here we assume that the variation vanishes at the end points of the worldline).
The energy-momentum tensor density $\sqrt{-g} T^{ab}$ is simply given by the coefficient
in front of $\delta g_{\mu\nu} / 2$ in (\ref{LMvar}). However, an explicit determination
requires yet another specialization of $\phi_I^{\text{field}}$, because the fields can
dependent on the metric. In the absence of $\phi_I^{\text{field}}$, one immediately
recovers the result of Tulczyjew \cite{Tulczyjew:1959}.
% In the next section we consider the simple example where $\phi_I^{\text{field}}$ is
% the curvature tensor.

\subsection{Quadrupole}
Let us now explore the case that $\phi_I^{\text{field}} = \{ R_{\mu\nu\alpha\beta} \}$.
It is useful to introduce an abbreviation for the corresponding partial derivative
of the Lagrangian,
\begin{equation}\label{Jmoment}
J^{\mu\nu\alpha\beta} := - 6 \frac{\partial L_M}{\partial R_{\mu\nu\alpha\beta}} .
\end{equation}
The conventional factor of $-6$ is motivated by comparing (\ref{pEOM}),
now reading
\begin{equation}\label{pEOMquad}
\frac{D p_{\mu} }{d \lambda} =
        \frac{1}{2} S_{\alpha\beta} R^{\alpha\beta}{}_{\rho\mu} u^{\rho}
        - \frac{1}{6} \nabla_{\mu} R_{\nu\rho\beta\alpha} J^{\nu\rho\beta\alpha} ,
\end{equation}
with the corresponding result of Dixon at the quadrupolar approximation level. An
identification of $J^{\mu\nu\alpha\beta}$ with Dixon's reduced quadrupole moment
is tempting, as this makes (\ref{pEOMquad}) formally identical to Dixon's result.
It is important that $J^{\mu\nu\alpha\beta}$ inherits the symmetries of
the Riemann tensor. From these symmetries and the properties of the operator
$G^{\mu}{}_{\nu}$, we obtain
% \begin{equation}
% G^{\nu}{}_{\mu} R_{\rho\sigma\alpha\beta}
% = - \delta^{\nu}_{\rho} R_{\mu\sigma\alpha\beta} - \delta^{\nu}_{\sigma} R_{\rho\mu\alpha\beta}
% - \delta^{\nu}_{\alpha} R_{\rho\sigma\mu\beta} - \delta^{\nu}_{\beta} R_{\rho\sigma\alpha\mu}
% \end{equation}
\begin{equation}\label{JGR}
J^{\sigma\rho\alpha\beta} G^{\mu}{}_{\nu} R_{\sigma\rho\alpha\beta}
= - 4 J^{\sigma\rho\alpha\beta} \delta^{\mu}_{\sigma} R_{\nu\rho\alpha\beta}
= - 4 J^{\mu\rho\alpha\beta} R_{\nu\rho\alpha\beta} .
\end{equation}
This simplifies (\ref{SEOM}) to
\begin{equation}
\frac{D S_{\mu\nu}}{d \lambda}
	= 2 p_{[\mu} u_{\nu]} + \frac{4}{3} R_{\alpha\beta\rho[\mu} J_{\nu]}{}^{\rho\beta\alpha} ,
\end{equation}
and formally agrees with Dixon's spin equation of motion, too.
Finally, the energy-momentum tensor agrees with the explicit
result in \cite{Steinhoff:Puetzfeld:2009} (in the present conventions, see
(5.3) in \cite{Steinhoff:2011}). This derives from
\begin{equation}
\delta R^{\mu}{}_{\nu\alpha\beta} = \nabla_{\alpha} \delta \Gamma^{\mu}{}_{\nu\beta}
	- \nabla_{\beta} \delta \Gamma^{\mu}{}_{\nu\alpha} , \label{Rvar}
\end{equation}
which must be further expanded using (\ref{Gvar}) and then leads to
\begin{equation}
\frac{\partial L_M}{\partial R_{\mu\nu\alpha\beta}} \delta_z R_{\mu\nu\alpha\beta}
= 	\int d^4 x \bigg[
 			- \frac{1}{3} J^{\mu\rho\alpha\beta} R^{\nu}{}_{\rho\alpha\beta} \delta_{(4)}
		- \frac{2}{3} \nabla_{\beta} \nabla_{\alpha} ( J^{\mu\alpha\beta\nu} \delta_{(4)} )
	\bigg] \frac{\delta g_{\mu\nu}}{2} .
\end{equation}
This is the contribution coming from the first term in (\ref{phidecompose}).
Another contribution arises from the second term in the first row of (\ref{LMvar}),
which is evaluated using (\ref{JGR}).
Collecting all terms in front of $\delta g_{\mu\nu}$ in (\ref{LMvar}), we can read
off the energy momentum tensor density as
\begin{equation}\label{EMT}
\begin{split}
\sqrt{-g} T^{\mu\nu} &= \int d \lambda \bigg[
	u^{(\mu} p^{\nu)} \delta_{(4)}
	- \nabla_{\alpha} ( S^{\alpha(\mu} u^{\nu)} \delta_{(4)} ) \nl
	+ \frac{1}{3} R_{\beta\alpha\rho}{}^{(\mu} J^{\nu)\rho\alpha\beta} \delta_{(4)}
        - \frac{2}{3} \nabla_{\beta} \nabla_{\alpha} ( J^{\mu(\alpha\beta)\nu} \delta_{(4)} )
\bigg] .
\end{split}
\end{equation}

\subsection{Other multipoles\label{multipoles}}
Dixon's moments are essentially defined as integrals over the energy-momentum tensor
of the extended body. Though these definitions can be applied to self-gravitating
bodies, the derivation of the equations of motions based on these definitions
only succeeds for test-bodies \cite{Dixon:1979}. It was shown in
\cite{Harte:2012} (see also the corresponding contribution by A. Harte in these proceedings)
using methods for self-force calculations that for
self-gravitating objects the equations of motions are still of the same form,
but the multipole moments must be renormalized.
The multipoles arising from the effective action should therefore be related
to these renormalized moments. For self-gravitating bodies, one can not in
general calculate the moments in the equations of motion using Dixon's integral formulas any more.
In the language of effective field theory, the multipoles are calculated through
a ``matching'' procedure instead, which will be explained in Sec.\ \ref{spinquad}.

Other gravitational multipoles can be incorporated by including symmetrized covariant derivatives
of the curvature in $\phi_I^{\text{field}}$. Similarly, electromagnetic multipoles arise from an analogous
construction based on the Faraday tensor $F_{\mu\nu}$. A quite exhaustive case
is therefore
\begin{equation}\label{generalfields}
\phi_I^{\text{field}} = \{ R_{\mu\nu\alpha\beta}, \nabla_{\rho} R_{\mu\nu\alpha\beta},
\nabla_{(\sigma} \nabla_{\rho)} R_{\mu\nu\alpha\beta}, \dots,
F_{\mu\nu}, \nabla_{(\rho} F_{\mu)\nu}, \nabla_{(\sigma} \nabla_{\rho} F_{\mu)\nu}, \dots \} .
\end{equation}
Notice that the commutation of covariant derivatives results in curvature terms
and that, e.g., $3 \nabla_{\alpha} F_{\mu\nu} =
2 \nabla_{(\alpha} F_{\mu)\nu} - 2 \nabla_{(\alpha} F_{\nu)\mu}$, which can
be checked using $\nabla_{[\alpha} F_{\mu\nu]} = 0$.
Again the partial derivatives of $L_M$ with respect to the $\phi_I^{\text{field}}$
can be called multipole moments. However, these multipoles and also
$p_{\mu}$ are probably not unique, because $L_M$ in not unique. For instance,
contractions of covariant derivatives with $u^{\mu}$ can be written as
$\lambda$-derivatives and one can partially integrate them. Notice that
Dixon's multipole moments have the same symmetries as ours, but satisfy additional
orthogonality relations to a timelike vector defined on the worldline.

It is also possible to include a term proportional to $A_{\mu} u^{\mu}$ in
the Lagrangian, as this combination transforms into a total $\lambda$-derivative
under a gauge transformation of the electromagnetic potential $A_{\mu}$. It just
leads to the well-known Lorentz force. However, in the present approach a
part of the Lorentz force is hidden in the definition of $p_{\mu}$,
making the equations of motion not manifestly gauge-invariant.

\section{Symmetries, transformations, and conditions\label{symetc}}
In this section we discuss symmetries, conservations laws, various transformations of the action,
and conditions it must fulfill.

\subsection{Symmetries and conserved quantities\label{conserved}}
Action principles have the advantage that one can easily derive conserved quantities
from the Noether theorem \cite{Noether:1918}.
Here we are going to consider only symmetry transformations where the fields are not
transformed. Further, we assume $\delta \lambda = 0$,
% (for implications of reparametrization invariance see, e.g., \cite{Steinhoff:2011}),
so the variational formula (\ref{LMvar}) together with (\ref{phidecompose}) is still valid.

On the one hand, we require
that the Lagrangian transforms under such a symmetry into a total derivative
\begin{equation}
\delta L_M = \frac{d K}{d \lambda} , \label{symcond}
\end{equation}
without making use of the equations of motion.
On the other hand, if we assume that the
equations of motion hold, then only the total time derivatives from (\ref{LMvar})
with (\ref{phidecompose}) inserted contribute to $\delta L_M$. These total derivatives
are located in the last lines of (\ref{LMvar}) and (\ref{phidecompose}).
(The first lines of (\ref{LMvar}) and (\ref{phidecompose}) vanishes because fields
are not transformed here.) We therefore have the conservation law
\begin{equation}\label{noether}
\frac{d}{d \lambda} \bigg[ p_{\mu} \delta z^{\mu}
        + \frac{1}{2} S_{\mu\nu} \Delta \Theta^{\mu\nu}
        + \psi^I_{\text{wl}} \delta \phi_I^{\text{wl}}
        - K \bigg] = 0 .
\end{equation}

A simple example is given by the global symmetry under a change of the
body-fixed frame. In order to make things even more simple, we assume
that $L_M$ does not depend on $\Lambda_A{}^{\mu}$ and on the $\dot{\phi}_I^{\text{wl}}$,
so that $\psi^I_{\text{wl}} = 0$.
But $L_M$ still implicitly depends on $\Lambda_A{}^{\mu}$
through $\Omega^{\mu\nu}$. A constant infinitesimal Lorentz transformation of
the body-fixed frame then reads
\begin{equation}
\delta z^a = 0 , \qquad
\delta \Lambda^{A\mu} = \omega^{AB} \Lambda_B{}^{\mu} ,
\end{equation}
where $\omega^{AB}$ is a constant infinitesimal antisymmetric matrix.
Obviously $\Omega^{\mu\nu}$ is invariant under this transformation, so
(\ref{symcond}) is fulfilled with $K = 0$. Further, we have
$\Delta \Theta^{ab} = \Lambda_A{}^a \Lambda_B{}^b \omega^{AB}$ and (\ref{noether}) reads
\begin{equation}
\frac{1}{2} \omega^{AB} \frac{d}{d \lambda} \bigg[ S_{\mu\nu} \Lambda_A{}^{\mu} \Lambda_B{}^{\nu} \bigg]
= 0 .
\end{equation}
As $\omega^{AB}$ is arbitrary, we see that the components of the spin in the
body-fixed frame $S_{AB} \equiv S_{\mu\nu} \Lambda_A{}^{\mu} \Lambda_B{}^{\nu}$ are constant.
A corollary of this fact is that the spin length $S$ is constant,
where $2 S^2 = S_{AB} S^{AB} = S_{\mu\nu} S^{\mu\nu}$.

The next important example is a symmetry of the spacetime
described by a Killing vector field $\xi^{\mu}$, $\mathcal{L}_{\xi} g_{\mu\nu} = 0$.
(Notice that also $\mathcal{L}_{\xi} R_{abcd} = 0$ etc.)
Other fields entering $L_M$ are assumed to be invariant under this symmetry, too, e.g.,
$\mathcal{L}_{\xi} F_{\mu\nu} = 0$.
We consider an infinitesimal shift of the worldline coordinate
\begin{equation}\label{wlshift}
\delta z^{\mu} = \epsilon \xi^{\mu} , \qquad
\Delta \Lambda_A{}^{\nu} = - \epsilon \mathcal{L}_{\xi} \Lambda_A{}^{\nu} = \epsilon \Lambda_A{}^{\mu} \nabla_{\mu} \xi^{\nu} , \qquad
\delta \phi_I^{\text{wl}} = 0 ,
\end{equation}
where $\epsilon$ is an infinitesimal constant and we assume parallel transport
of $\Lambda_A{}^{\mu}$ along $\xi^{\nu}$, i.e.,
$\xi^{\nu} \nabla_{\nu} \Lambda_A{}^{\mu} = 0$.
Notice that the fields are not transformed, but their symmetry along $\xi^{\mu}$ is important.
Recall that $- \epsilon \mathcal{L}_{\xi}$ generates an infinitesimal coordinate transformation.
Therefore, $L_M$ is invariant under this transformation if all the variables it
depends on, including the fields, would be transformed by $- \epsilon \mathcal{L}_{\xi}$. But the shift
(\ref{wlshift}) only applies to $z^{\mu}$, $\Lambda_A{}^{\mu}$, and $\phi_I^{\text{wl}}$, so the
result of (\ref{wlshift}) on $L_M$ is exactly opposite to the case when all
variables \emph{except} $z^{\mu}$, $\Lambda_A{}^{\mu}$, and $\phi_I^{\text{wl}}$ are transformed. These variables
are all the fields, so the $\delta L_M$ produced by (\ref{wlshift}) can be obtained by
transforming all the fields using $+ \epsilon \mathcal{L}_{\xi}$. But the fields were assumed to be invariant
under this transformation. Hence we have argued that (\ref{wlshift}) is a symmetry of
the action, $\delta L_M=0$, and $K=0$. Combining (\ref{noether}) and (\ref{wlshift}),
we find the conserved quantity
\begin{equation}\label{Killconst}
E_{\xi} := p_{\mu} \xi^{\mu} + \frac{1}{2} S^{\mu\nu} \nabla_{\mu} \xi_{\nu} = \text{const} ,
\end{equation}
where $\Delta \Theta^{\mu\nu} = \epsilon \nabla^{[\mu} \xi^{\nu]}$ was used. It is
interesting that this covers to all multipole orders. This was also shown in
\cite[p.\ 210]{Ehlers:Rudolph:1977} based on the equations of motion.

A special kind of conserved quantities that is not covered by the Noether theorem here
are mass-like quantities. We will see later on that masses enter the action as
parameters and are therefore constant by assumption.

\subsection{Legendre transformations\label{problems}}
Before proceeding, it is worth to point out that of course not every conceivable
Lagrangian $L_M$ is acceptable. Some choices are mathematically inconsistent or
physically unacceptable for other reasons. Some Lagrangians $L_M$ are technically more difficult to
handle and it makes sense to assume simplifying conditions for $L_M$ for a first study.
One such assumption we make here is that the relation between spin and angular velocity
is a bijection. Notice that this relation is fixed by $L_M$ through our definition of
$S_{\mu\nu}$ in (\ref{Lder}). A violation of our assumption can have the interesting implication
that the spin supplementary condition follows from (\ref{Lder}), see \cite{Hanson:Regge:1974},
but we will not considering this scenario here.
The supplementary conditions are discussed in the next section.

With this assumption on the relation between spin and angular velocity, we can
solve for $\Omega^{\mu\nu}$ in terms of $S^{\mu\nu}$ (and probably other variables).
This allows a Legendre transformation in $\Omega^{\mu\nu}$, i.e.,
\begin{equation}
W_M[e_a{}^{\mu}, z^{\rho}, \Lambda_A{}^{\mu}, S_{\mu\nu}, \dots] = \int d \lambda \left[
        \frac{1}{2} S_{\mu\nu} \Omega^{\mu\nu}
	+ R_M(g_{\mu\nu}, u^{\mu}, \Lambda_A{}^{\mu}, S_{\mu\nu}, \phi_I) \right] .
\end{equation}
It is important that the spin is varied independently now. Notice that this notion
of Legendre transformation is unusual in mechanics, as $\Omega^{\mu\nu}$ is not a time
derivative, but a combination of time derivatives. Still Legendre transformations are
applicable in much more generic situations, which is heavily used, i.e., in thermodynamics.
The function $R_M$ establishes the connection to Routhian approaches
\cite{Yee:Bander:1993, Porto:Rothstein:2008:1, Porto:Rothstein:2008:2}.
The Routhian is a mixture of a Hamiltonian
and a Lagrangian. Here it is essentially the sum of
$R_M$ and the connection part in $\frac{1}{2} S_{\mu\nu} \Omega^{\mu\nu}$. Notice that
therefore the Routhian is not manifestly covariant and covariance only becomes apparent
at the level of the equations of motion. In contrast, in our construction $R_M$ is
manifestly covariant.

A consequence of reparametrization invariance is that $L_M$ must be a homogeneous function of
degree one in all (first-order) $\lambda$-derivatives. For our assumptions
in Sec.\ \ref{EOMassumptions} this applies
only to $u^{\mu}$, $\Omega^{\mu\nu}$, and $\dot{\phi}_I^{\text{wl}}$,
so Eulers theorem on homogeneous functions reads
\begin{equation}\label{reparaID}
L_M = \frac{\partial L_M}{\partial u^{\mu}} u^{\mu}
        + \frac{\partial L_M}{\partial \Omega^{\mu\nu}} \Omega^{\mu\nu}
        + \frac{\partial L_M}{\partial \dot{\phi}_I^{\text{wl}}} \dot{\phi}_I^{\text{wl}}
= p_{\mu} u^{\mu} + \frac{1}{2} S_{\mu\nu} \Omega^{\mu\nu}
        + \psi^I_{\text{wl}} \dot{\phi}_I^{\text{wl}} .
\end{equation}
This is a consequence of reparametrization-gauge invariance, so in this sense
it is analogous to (\ref{LMidentity}), which follows from coordinate-gauge invariance.
Let us proceed with a Legendre transformation in $u^{\mu}$. This is more subtle, as the relation
between $u^{\mu}$ and $p_{\mu}$ can not be a bijection.
To see this, first notice that (\ref{reparaID}) can be interpreted as a constraint
on the component $p_{\mu} u^{\mu}$ of $p_{\mu}$.
This can be formulated as the famous mass-shell constraint
\begin{equation}\label{massshell}
p_{\mu} p^{\mu} + \mathcal{M}^2 = 0 ,
\end{equation}
where $\mathcal{M}$ is called dynamical mass and usually depends on the dynamical
variables.
Thus the momentum only has three independent components.
On the other hand, $u^{\mu}$ has four independent components:
three physical and one gauge degree of freedom due to reparametrization invariance
in $\lambda$.
(If we would choose $\lambda$ to be the proper time, then
just 3 components are independent. But the constraint $u^{\mu} u_{\mu} = -1$
makes the variational principle more subtle.) That is, the constraint (\ref{massshell})
produces a mismatch in degrees of freedom between $u^{\mu}$ and $p_{\mu}$, so they can
not be connected by a bijection. However, the Legendre
transformation can in fact be generalized to the case where constraints appear. One can
perform the Legendre transformation ``as usual'' if all constraints are added to the action
using Lagrange multipliers \cite{Rosenfeld:1930, Dirac:1950}. Here we need one Lagrange
multiplier $\alpha$ for (\ref{massshell}), which together with the three independent
components of $p_{\mu}$ provides a total of four independent
variables. This exactly matches the four independent degrees of freedom of $u^{\mu}$.
The Lagrange multiplier $\alpha$ isolates the reparametrization-gauge degree of freedom,
while $p_{\mu}$ represents the physical degrees of freedom.
Again we require $L_M$ to be such that no pathologies
for this ``constraint'' Legendre transformation arise.

Finally, the result of the transformation is
\begin{multline}
W_M[e_a{}^{\mu}, z^{\rho}, p_{\mu}, \alpha, S_{\mu\nu}, \Lambda_A{}^{\mu},
        \phi_I^{\text{wl}}, \psi^I_{\text{wl}}, \dots] \\
        = \int d \lambda \bigg[
        p_{\mu} u^{\mu}
        + \frac{1}{2} S_{\mu\nu} \Omega^{\mu\nu}
        + \psi^I_{\text{wl}} \dot{\phi}_I^{\text{wl}}
	- \frac{\alpha}{2} ( p_{\mu} p^{\mu} + \mathcal{M}^2 ) \bigg] ,
\end{multline}
where
\begin{equation}
\mathcal{M} = \mathcal{M}(g_{\mu\nu}, p_{\mu}, \Lambda_A{}^{\mu}, S_{\mu\nu},
        \phi_I^{\text{wl}}, \psi^I_{\text{wl}}, \phi_I^{\text{field}}) .
\end{equation}
We assume that we can also Legendre transform in the $\dot{\phi}_I^{\text{wl}}$
without giving rise to further constraints or pathologies.
From the variations of $p_{\mu}, S_{\mu\nu}$, and $\psi^I_{\text{wl}}$, we obtain
\begin{equation}\label{uLegendre}
u^{\mu} = \alpha p^{\mu} + \frac{\alpha}{2} \frac{\partial \mathcal{M}^2}{\partial p_{\mu}} , \qquad
\Omega^{\mu\nu} = \alpha \frac{\partial \mathcal{M}^2}{\partial S_{\mu\nu}} , \qquad
\dot{\phi}_I^{\text{wl}} = \frac{\alpha}{2} \frac{\partial \mathcal{M}^2}{\partial \psi^I_{\text{wl}}} .
\end{equation}
These are just the inverses to variable transformations used in the Legendre transformations.
Because we did not touch the variables $g_{\mu\nu}$, $\phi_I^{\text{wl}}$,
$\Lambda_A{}^{\mu}$ and $\phi_I^{\text{field}}$,
it is clear that
\begin{align}\label{derLM}
\frac{\partial L_M}{\partial g_{\mu\nu}} &\equiv
        - \frac{\alpha}{2} \frac{\partial \mathcal{M}^2}{\partial g_{\mu\nu}}, &
\frac{\partial L_M}{\partial \Lambda_A{}^{\mu}} &\equiv
        - \frac{\alpha}{2} \frac{\partial \mathcal{M}^2}{\partial \Lambda_A{}^{\mu}}, \\
\frac{\partial L_M}{\partial \phi_I^{\text{wl}}} &\equiv
        - \frac{\alpha}{2} \frac{\partial \mathcal{M}^2}{\partial \phi_I^{\text{wl}}}, &
\frac{\partial L_M}{\partial \phi_I^{\text{field}}} &\equiv
        - \frac{\alpha}{2} \frac{\partial \mathcal{M}^2}{\partial \phi_I^{\text{field}}}.
        \label{LMwl}
\end{align}
The Lagrange multiplier $\alpha$ is determined by choosing a normalization for $u^{\mu}$,
which corresponds to a gauge choice for $\lambda$.
For a given dynamical mass function $\mathcal{M}$, one can then evaluate the equations of motion
(\ref{pEOM}) and (\ref{SEOM}).

Coming back to the plan outlined in the introduction, we have the option to
construct either $L_M$, $R_M$, or $\mathcal{M}$ in a  phenomenological manner.
Let us explore the last option here, e.g., because
it promotes both $p_{\mu}$ and $S_{\mu\nu}$ to dynamical variables, which are probably easier to identify in
realistic situations compared to $u^{\mu}$ and $\Omega^{\mu\nu}$.
% This choice is also in accord with ideas from \cite[Sec.\ VI.A]{Blanchet:Buonanno:LeTiec:2012}.
Further, it is suggestive that the mass $\mathcal{M}$
of the object as a function of the dynamical variables completely determines the macroscopic dynamics of
the body. This situation is analogous to a thermodynamic potential (like the internal energy)
describing the large-scale behavior of a thermodynamic system. This is the first indication
that thinking in terms of thermodynamic analogies is very useful here.
% Furthermore, $\mathcal{M}$ can be adopted to certain famous spin supplementary conditions very well.

\subsection{Supplementary conditions}
The model for spinning bodies developed up to now comprises too many degrees of freedom.
We expect three rotational degrees of freedom instead of six provided by the
Lorentz frame $\Lambda_A{}^{\mu}$. Similarly, the spin should only have three independent
components, too. It is suggestive to impose that the time direction of the body-fixed
frame is aligned to a (to be defined) rest frame described by a unit time-like vector $r^{\mu}$,
and that the spin only has spatial components in this rest frame,
\begin{equation}\label{conditions}
\Lambda_0{}^{\mu} = r^{\mu} , \qquad
S_{\mu\nu} r^{\nu} = 0 .
\end{equation}
One can also envision different time-like vectors in each of these conditions. However, using
the same vector seems to fit well to the interpretation of $r^{\mu}$ as a rest frame.
The condition on the spin is usually called spin supplementary condition. Two
specific options are $r^{\mu} = u^{\mu} / \sqrt{-u_{\nu}u^{\nu}}$ or
$r^{\mu} = p^{\mu} / \sqrt{-p_{\nu}p^{\nu}}$. The latter condition is usually
considered as the best choice, as it uniquely fixes the representative worldline of the extended
object \cite{Beiglbock:1967, Schattner:1979:1, Schattner:1979:2} (if Dixon's definitions
for the multipoles are used).
A more detailed discussion of supplementary conditions is given in the contributions by
D.~Giulini, L.F.~Costa and J.~Nat{\'a}rio. But notice
that in flat space the choice of this condition can be related to the choice of the representative
worldline for the extended body. In curved spacetime this relation could not be established
yet. From a careful perspective one should therefore reckon that different spin supplementary
conditions may lead to inequivalent models. As long as the relation to the choice of center is not
clarified for curved spacetimes, one must regard this condition as a constitutive relation of
the model. For this reason, one should also avoid conditions which are not manifestly covariant.

The most straightforward way to implement (\ref{conditions}) into a given action is to add
these conditions using Lagrange multipliers. In general, this will modify the dynamics by
constraint forces. As in classical mechanics, one requires that (\ref{conditions}) is preserved
in time, which should fix the Lagrange multipliers. This can lead to inconsistencies, in which
case one should revise the action or the choice for $r^{\mu}$. It can also lead to further
constraints, which we regard as unphysical here as they further reduces the number of independent
variables (we want exactly three rotational degrees of freedom). Similarly, if some of the Lagrange
multipliers remain undetermined, then the degrees of freedom are increased, which we
also regard as unphysical. The last possibility is that the Lagrange multipliers are uniquely
fixed by requiring that (\ref{conditions}) is preserved. In the end, we can insert this solution
for the Lagrange multipliers into the action. In this way we obtain an action without Lagrange
multipliers which preserves (\ref{conditions}).

\subsection{Conditions on the dynamical mass\label{SSCcond}}
Having this said, we can try to directly construct an action which preserves (\ref{conditions}).
This approach is in fact very natural here.
For instance, one can make an ansatz for $\mathcal{M}^2$ and use this requirement to fix
some of the coefficients. The first condition in (\ref{conditions}) can be written as
$\eta_{0A} = \Lambda_{A\mu} r^{\mu}$ and is preserved in time if
\begin{equation}\label{compat1}
0 = \frac{D r^{\mu}}{d \lambda} + \Omega^{\mu\nu} r_{\nu} , \qquad
\text{where } \Omega^{\mu\nu} = \alpha \frac{\partial \mathcal{M}^2}{\partial S_{\mu\nu}} .
\end{equation}
Using (\ref{LMidentity}) and (\ref{derLM}), we can write the spin equation of motion (\ref{SEOM}) in the form
\begin{align}
\frac{D S^{\mu\nu}}{d \lambda} &= 2 S_{\alpha}{}^{[\mu} \Omega^{\nu]\alpha}
        + \alpha \frac{\partial \mathcal{M}^2}{\partial \Lambda_A{}^{\alpha}} \delta_{\alpha}^{[\mu} \Lambda_A{}^{\nu]} .
\end{align}
With the help of this equation, we see that the spin supplementary condition is preserved in
time if it holds (\ref{compat1}) and additionally
\begin{equation}\label{compat2}
0 = \frac{\partial \mathcal{M}^2}{\partial \Lambda_A{}^{\alpha}} \delta_{\alpha}^{[\mu} \Lambda_A{}^{\nu]} r_{\nu} .
\end{equation}
This condition is often trivially fulfilled, namely when $\mathcal{M}^2$ does not
explicitly depend on $\Lambda_A{}^{\mu}$.
We are going to construct a simple example now, in order to show that functions $\mathcal{M}$
exist which are consistent with all of our requirements.

\subsection{A simple construction of the dynamical mass}
Instead of constructing an action which fulfills (\ref{compat1}) and (\ref{compat2}) for
a specific choice of $r^{\mu}$, one can
look at a specific action and construct a $r^{\mu}$ such that the requirements (\ref{compat1})
and (\ref{compat2}) are fulfilled. Let us consider a simple example where $\mathcal{M}^2$ is
a nonconstant analytic function $f$ depending only on $S^2 := S^{\mu\nu} S_{\mu\nu} / 2$,
i.e., $\mathcal{M}^2 = f(S^2)$.
It is clear that (\ref{compat2}) is fulfilled, because there is no explicit
dependence on the body-fixed frame. We still need to satisfy (\ref{compat1}),
which reads explicitly
\begin{equation}\label{revolve}
0 = \frac{D r^{\mu}}{d \lambda} + \alpha f' S^{\mu\nu} r_{\nu}
= \frac{D r^{\mu}}{d \lambda} .
\end{equation}
That is, the vector $r^{\mu}$ must be parallel transported along the worldline.
This does indeed characterize a suitable spin supplementary condition, which
was first discussed in \cite[Sec.\ 3.4]{Kyrian:Semerak:2007}.
Although, with this condition, $r^{\mu}$ lacks an immediate interpretation as a rest frame,
the numerical results in \cite{Kyrian:Semerak:2007} show that it leads to
similar predictions as the choice $r^{\mu} = p^{\mu} / \sqrt{-p_{\nu}p^{\nu}}$.
Further discussions on this supplementary condition are given in
other contributions to these proceedings.
% Though $r^{\mu} = p^{\mu} / \sqrt{-p_{\nu}p^{\nu}}$ is more difficult to arrange, we will restrict to this case from now on.

For the case of a black hole, the laws of black hole dynamics
\cite[Box 33.4]{Misner:Thorne:Wheeler:1973} suggests that
\begin{equation}
\mathcal{M}^2_{\text{BH}} = f(S^2) = m_0^2 + \frac{S^2}{(2 G m_0)^2} .
\end{equation}
where $m_0$ is the constant irreducible mass related to the horizon area.
We can now have a look at the angular velocity with respect to asymptotic time,
so we have $\alpha = \mathcal{M}^{-1}$ (for a body at rest).
Evaluating (\ref{uLegendre}), we find agreement
with what is usually identified as the angular velocity of the horizon. This is
a nice check for the consistency of the interpretation of our variables.
Notice that the laws of black hole dynamics
owe their name to their similarity to the laws of thermodynamics.
Further, an action principle similar to the one presented here can
be used to derive the so called first law of black hole \emph{binary} dynamics
\cite{Blanchet:Buonanno:LeTiec:2012}. Again we encounter the thermodynamic
character of the approach.

For objects other than black holes, we can derive $\mathcal{M}$ from the
moment of inertia. One usually defines the moment of inertia $I(S^2)$
as the proportionality factor between spin and angular velocity, which can be
read off from (\ref{uLegendre}).
Again we have $\alpha = \mathcal{M}^{-1} = f^{-1/2}$, so (\ref{uLegendre})
leads to the differential equation $I^{-1} = f^{-1/2} f'$.
Its solution reads
\begin{equation}\label{simpleM}
\mathcal{M}^2 = f(S^2) = \left[ m_0 + \int_0^{S^2} \frac{d x}{2 I(x)} \right]^2 ,
\end{equation}
where the irreducible mass $m_0$ enters as an integration constant.
For neutron stars, the function $I(S^2)$ can be obtained numerically, e.g., using the RNS code
\cite{Stergioulas:Morsink:1999, Stergioulas:Friedman:1995}.
Alternatively, one can numerically compute the gravitating mass $\mathcal{M}$ directly
as a function of $S$ for a fixed number of baryons in the star.
It would be interesting to see if both methods lead to compatible results.
It should be noted that both black holes and neutron stars posses a quadrupole
(and other multipoles) when they are spinning, which was neglected here.
It will be included in the next section.

Interestingly, it is implied by \cite{Kunzle:1972} that for the pole-dipole case
one can construct a $\mathcal{M}^2$ such that (\ref{compat1}) and (\ref{compat2}) are
fulfilled for $r^{\mu} = p^{\mu} / \sqrt{-p_{\nu}p^{\nu}}$ without the need for
approximations or truncations of $\mathcal{M}^2$. Then, however, $\mathcal{M}^2$ is
not solely dependent on $S$. The details on this are left for a future work.

\section{Spin-induced quadrupole\label{spinquad}}
In this section, we are going to develop a simple phenomenological model for
$\mathcal{M}^2$ describing
the spin-induced quadrupole of a star. This is the quadrupole of a star
arising from a deformation away from spherical symmetry due to rotation.
We start with a reasonable ansatz for $\mathcal{M}^2$.
The main idea for this ansatz is to include all possible covariant
(general coordinate invariant) terms up to a certain power in spin and curvature.
The unknown coefficients
in this ansatz are then fixed by comparing to the Kerr metric and to
numerical solutions for the gravitational field of a rotating neutron star.
One should emphasize that a truncation of $\mathcal{M}^2$ requires negligibly small interaction
energies, not small multipoles.

\subsection{Construction of the action\label{construct}}
We are going to include in our ansatz the quadratic order in spin and second order
derivatives of the metric. This means that we include terms linear in the curvature
and covariant derivatives of the curvature are not allowed. 
This implies that we exclude $\lambda$-derivatives of the curvature for now, but
this restriction will be loosened below.
Symbolically we have $\phi_I^{\text{field}} = \{ R_{\mu\nu\alpha\beta} \}$,
which according to Sec.\ \ref{multipoles} implies that we
neglect interaction terms involving octupole and higher multipoles.
Finally, let us assume the absence of further worldline degrees of freedom in this section,
or $\phi_I^{\text{wl}} = \emptyset$, so we have no need for a dependence of
$\mathcal{M}^2$ on $\Lambda_A{}^{\mu}$. [Then (\ref{compat2}) is already fulfilled.]

The main task is to collect all possible interaction terms. One must
take care of including only independent terms, which can by tricky due
to the symmetries of the Riemann tensor.
A procedure for this was applied to the construction of
effective Lagrangians or Routhians in
\cite{Porto:Rothstein:2008:2, Goldberger:Rothstein:2006,
Goldberger:Rothstein:2006:2}. Instead, we are going to construct $\mathcal{M}^2$ directly,
but the arguments are essentially the same.
We will follow a different approach to implement the spin supplementary condition, too,
by making an ansatz for $r^{\mu}$ around the case
\begin{equation}\label{rpre}
r^{\mu} = \frac{p^{\mu}}{\sqrt{-p_{\nu}p^{\nu}}} + \Order(R_{\mu\nu\alpha\beta}).
\end{equation}

As a first simplification, one can replace $R_{\mu\nu\alpha\beta}$ by its tracefree
version, the Weyl tensor $C_{\mu\nu\alpha\beta}$. The traces are given by
$R_{\mu\nu} := g^{\alpha\beta} R_{\mu\alpha\nu\beta}$ and $R := g^{\mu\nu} R_{\mu\nu}$,
which are related to the energy momentum tensor $T^{\mu\nu}$ through
Einstein's gravitational field equations
\begin{equation}\label{EFE}
R^{\mu\nu} = 8\pi G \left( T^{\mu\nu} -\frac{1}{2} T^{\alpha}{}_{\alpha} g_{\mu\nu} \right).
\end{equation}
The energy momentum tensor can contain contributions from fields penetrating the
compact object, like electromagnetic or dark matter fields. We assume that these
can be neglected, i.e., the bodies are mainly interacting via the gravitational field.
But in the case of self-gravitating bodies, the energy momentum tensor also includes
a singular contribution from the point-particle (\ref{EMT}) itself. Let us assume
that these singular self-interactions can be dropped. Then we can effectively
make use of the vacuum field equations $R^{\mu\nu} = 0$ at the particle location,
so we have $R_{\mu\nu\alpha\beta} = C_{\mu\nu\alpha\beta}$.
However, in general one is not allowed to use field equations at the level of the action.
But in the current context this is essentially a valid procedure, as it is equivalent
to a field redefinition in the action, see \cite{Goldberger:Rothstein:2006},
\cite{Damour:Schafer:1991}, or \cite[Appendix A]{Damour:EspositoFarese:1998}.
Without loss of generality, we can therefore restrict to
$\phi_I^{\text{field}} = \{ C_{\mu\nu\alpha\beta} \}$ in the quadrupole case.

The most important and most obvious requirement on the allowed interaction terms
in $\mathcal{M}^2$ is general coordinate invariance. Further restrictions on the
terms and transformations identifying equivalent terms (equivalent within our truncation)
are:
\begin{enumerate}
\item In four spacetime dimensions, the Weyl tensor can be split into an electric
$E_{\mu\nu}$ and a magnetic part
$B_{\mu\nu}$ with respect to a time-like unit vector. Choosing this vector to be
$r^{\mu}$, it holds
\begin{equation}\label{defEB}
E_{\mu\nu} = C_{\mu\alpha\nu\beta} r^{\alpha} r^{\beta} , \qquad
B_{\mu\nu} = \frac{1}{2} \eta_{\mu\alpha\rho\sigma} C_{\nu\beta}{}^{\rho\sigma} r^{\alpha} r^{\beta} ,
\end{equation}
where $\eta_{\mu\nu\alpha\beta}$ is the volume form. These tensors have the properties
\begin{align}
E_{\mu\nu} &= E_{\nu\mu} , &
E_{\mu\nu} g^{\mu\nu} &= 0 , &
E_{\mu\nu} r^{\nu} &= 0 , \\
B_{\mu\nu} &= B_{\nu\mu} , &
B_{\mu\nu} g^{\mu\nu} &= 0 , &
B_{\mu\nu} r^{\nu} &= 0 .
\end{align}
These properties make $E_{\mu\nu}$ and $B_{\mu\nu}$ much easier to handle compared
to $C_{\mu\nu\alpha\beta}$.
\item We include only terms invariant under parity transformations. In this respect
it is important to notice that $B_{\mu\nu}$ is of odd parity.
We conclude that any terms with an odd number of magnetic Weyl tensors must
also include an odd number of volume forms $\eta_{\mu\nu\alpha\beta}$.
Due to the antisymmetry of $\eta_{\mu\nu\alpha\beta}$, it
will always be contracted with both indices of the spin at the current level of truncation.
Then we can rewrite all terms involving $\eta_{\mu\nu\alpha\beta}$ in terms of the dual of
the spin tensor
\begin{equation}
*\!S_{\alpha\beta} := \frac{1}{2} S^{\mu\nu} \eta_{\mu\nu\alpha\beta} .
\end{equation}
Notice that we have the identity \cite[Eq.\ (A.7)]{Hanson:Regge:1974}
\begin{equation}\label{SSstar}
S_{\mu\alpha} *\!S^{\alpha\nu} = - \frac{1}{4} \delta_{\mu}^{\nu} S_{\alpha\beta} *\!S^{\alpha\beta} .
\end{equation}
As a consequence, it holds $B_{\mu\nu} S^{\mu}{}_{\alpha} *\!S^{\alpha\nu} = 0$.
It is customary to define a spin vector $S^{\mu} := r_{\nu} *\!S^{\nu\mu}$.
\item It should be noticed that one is in general not allowed to neglect terms
involving the combination $S_{\mu\nu} r^{\nu}$, though these numerically vanish
due to the spin supplementary condition (\ref{conditions}): A variation of these terms can lead to
nonvanishing contributions to the equations of motion. Instead, terms in the action
which are at least quadratic in $S_{\mu\nu} r^{\nu}$ can be neglected, as their contributions
to the equations of motion are at least linear in the spin supplementary condition and
thus always vanish.
\item The quadrupole interaction terms can be simplified using the leading order truncation
of the mass-shell constraint
$p_{\mu} p^{\mu} + \mathcal{M}^2 = 0$. (For the ansatz in (\ref{Mansatz}) given below, this implies that we can
set $p_{\mu} p^{\mu} \approx - \mu^2$ in the higher order terms of $\mathcal{M}^2$.)
This transformation does in fact just correspond to a redefinition of the Lagrange multiplicator
$\alpha$, and the idea is therefore similar to the field redefinitions mentioned above
\cite{Damour:Schafer:1991}.
\item Time derivatives of $p_{\mu}$ and $S_{\mu\nu}$ can also be removed by redefinitions
of variables, which follows from the ideas in \cite{Damour:Schafer:1991} and is again
analogous to the mentioned field redefinitions. Besides that, the absence of higher order
time derivatives was already assumed in Sec.\ \ref{EOMassumptions}.
\end{enumerate}

The last point also shows that our ansatz will automatically cover time derivatives of
$E_{\mu\nu}$ and $B_{\mu\nu}$ of arbitrary order. As we work at linear level in the curvature,
the time derivatives can always be removed from the curvature through partial integration.
After this transformation, all time derivatives finally apply to $p_{\mu}$ and $S_{\mu\nu}$ only,
which can be removed by virtue of the argument 5 above. This suggests
that these terms belong to the quadrupole level, too, although time derivatives
of the fields are in fact covariant derivatives $D / d \lambda = u^{\mu} \nabla_{\mu}$.
This is of course related to the ambiguity of the multipoles pointed out in
Sec.\ \ref{multipoles}. At linear level in the curvature, one can assume that the
covariant derivatives are projected orthogonal to $r^{\mu}$, because
% from (\ref{ransatz}) and (\ref{uLegendre}) we see that
$r^{\mu} \nabla_{\mu} \approx u^{\mu} \nabla_{\mu} = D / d \lambda $ up to higher order terms,
which can be partially integrated.

The first point mentioned above suggests to include just
$E_{\mu\nu}$ and $B_{\mu\nu}$ in $\phi_I^{\text{field}}$.
However, this is currently not possible, because we assumed in Sec.\ \ref{EOMassumptions} that the
$\phi_I^{\text{field}}$ contain
just fields, but $r^{\mu}$ in (\ref{defEB}) is only defined on the worldline. For instance,
one would have to clarify
the meaning of $\nabla_{\alpha} r^{\mu}$ arising from $\nabla_{\alpha} E_{\mu\nu}$ in (\ref{pEOM}).
For simplicity, let us stick to $\phi_I^{\text{field}} = \{ C_{\mu\nu\alpha\beta} \}$ here,
but have in mind that $\mathcal{M}^2$ depends on $C_{\mu\nu\alpha\beta}$ only through the
combinations $E_{\mu\nu}$ and $B_{\mu\nu}$.
The equations of motion are initially expressed in terms the quadrupole moment related to
$C_{\mu\nu\alpha\beta}$,
\begin{equation}\label{Jmoment2}
\tilde{J}^{\mu\nu\alpha\beta} := - 6 \frac{\partial L_M}{\partial C_{\mu\nu\alpha\beta}} .
\end{equation}
see (\ref{Jmoment}), but these are at once related to the moments belonging to
$E_{\mu\nu}$ and $B_{\mu\nu}$ through the chain rule.
The interpretation of the latter moments as
quadrupoles is much more obvious than for (\ref{Jmoment2}), as $E_{\mu\nu}$ and $B_{\mu\nu}$
are symmetric tracefree spatial tensors in the rest frame defined by $r^{\mu}$.
These moments can be called
electric and magnetic quadrupoles, respectively. They match the quadrupole degrees
of freedom of the gravitational field outside the body \cite{Thorne:1980}, in contrast
to (\ref{Jmoment}), which in general is not tracefree.
This approach to define electric and magnetic quadrupoles was briefly discussed
in \cite{Steinhoff:2011}. An explicit split into electric and magnetic quadrupoles
at the level of the equations of motion was performed in \cite{Bini:Geralico:2014}.

\subsection{Ansatz}
The most general ansatz for $\mathcal{M}^2$ now reads
\begin{equation}\label{Mansatz}
\mathcal{M}^2 = \mu^2
+ C_{BS^2p} B_{\mu\nu} S^{\mu} S^{\nu\alpha} \hat{p}_{\alpha}
+ C_{ES^2} E_{\mu\nu} S^{\mu\alpha} S^{\nu}{}_{\alpha}
+ \Order(E^2, B^2, S^3) ,
\end{equation}
where we introduced the abbreviation $\hat{p}_{\mu} = p_{\mu} / \mu$. We assume that
$\mu$, $C_{BS^2p}$, and $C_{ES^2}$ are constants. Remember that within the curvature
terms we can set $r^{\mu} \approx \hat{p}^{\mu}$, which is due to (\ref{rpre}) and point 4
of the last section.
Notice that $\mu$ must be a function of the constant spin length,
$\mu^2 = f(S^2)$, cf.\ (\ref{simpleM}). Otherwise the Legendre transformation would be problematic.
Consistent with our truncation, we may write
\begin{equation}
\mu^2 = m_0^2 + \frac{m_0}{I_0} S^2 + \Order(S^4),
\end{equation}
where $I_0 \equiv I(0)$ is the moment of inertia in a slow
rotation limit $S \rightarrow 0$.
Furthermore, the constants $C_{BS^2p}$ and $C_{ES^2}$ will in general depend on
$\mu$ and $S$. This is further discussed below.

Next, we want to check if (\ref{compat1}) is fulfilled. Notice that
(\ref{compat1}) is required to hold at linear order in spin only.
For this purpose, let us make an ansatz for $r^{\mu}$ to linear order in $S$,
\begin{equation}\label{ransatz}
r^{\mu} = C_{rp} \hat{p}^{\mu} + G C_{rBS} B^{\mu\nu} S_{\nu} + \Order(E^2, B^2, S^2) .
\end{equation}
The normalization $r^{\mu} r_{\mu} = -1$ leads to $C_{rp} = 1$.
Inspecting (\ref{compat1}), we see that most of the contributions
from the $C_{rBS}$-term are shifted to higher orders, namely quadratic level
in spin: This is due to $\Omega = \Order(S)$ and $\dot{S}_{\nu} = \Order(S^2)$.
The only $C_{rBS}$-term linear in spin contains $\dot{B}^{\mu\nu}$. Though this is a
derivative of the curvature, it is not of higher order, because we realized that
our ansatz effectively also covers $\lambda$-derivatives of the curvature.
We conclude that $C_{rBS} = 0$, or $r^{\mu} \approx \hat{p}^{\mu}$.
For calculating $\dot{r}^{\mu}$ in (\ref{compat1}), it is useful to rewrite
(\ref{pEOM}) as
\begin{equation}
\frac{D p_\mu}{d \lambda} = \alpha ( E_{\alpha[\mu} S_{\nu]}{}^{\alpha}
        - B_{\alpha[\mu} r_{\nu]} S^{\alpha} )
        \left[ 2 p^{\nu} + \frac{\partial \mathcal{M}^2}{\partial p_{\nu}} \right]
        - \frac{\alpha}{2} (\nabla_{\mu} \phi_I^{\text{field}}) \frac{\partial \mathcal{M}^2}{\partial \phi_I^{\text{field}}} .
\end{equation}
Finally, the condition (\ref{compat1}) is fulfilled to linear order in spin if
$C_{BS^2p} = 2$ in our ansatz (\ref{Mansatz}).
The condition (\ref{compat2}) is of course also fulfilled.
In summary, we must have
\begin{equation}
C_{rp} = 1, \qquad
C_{rBS} = 0, \qquad
C_{BS^2p} = 2 ,
\end{equation}
while $C_{ES^2}$ is not determined by basic principles, but depends on the specific object.
Instead of fixing $r^{\mu}$ algebraically like in (\ref{ransatz}), it would be
interesting to view (\ref{compat1}) as an evolution equation for $r^{\mu}$
in the future, analogous to (\ref{revolve}) in the pole-dipole case.

\begin{figure}
\begin{center}
\includegraphics{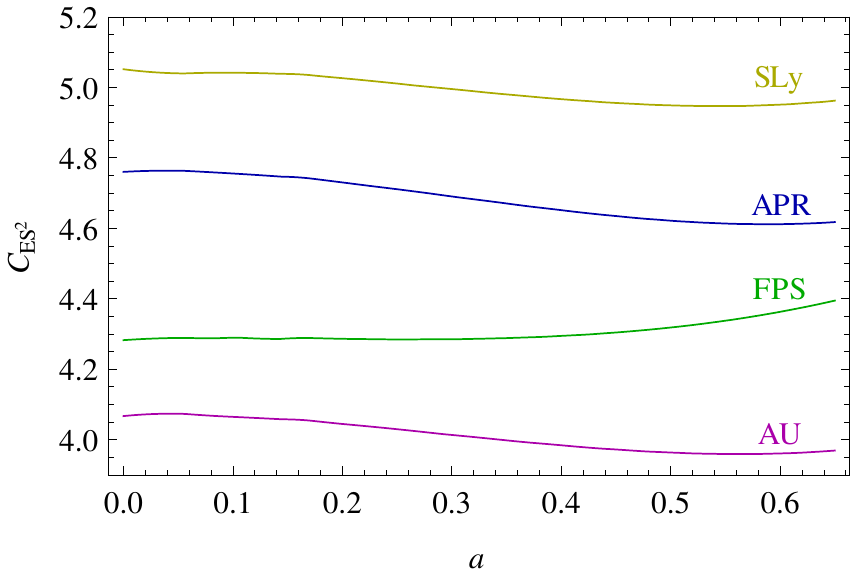}
\end{center}
\caption{The coefficient $C_{ES^2}$ as a function of the dimensionless spin
$a = S / G \mu^2$, where $\mu$ is identified with the gravitating mass
and is given by $\mu = 1.4 M_{\astrosun}$ here. The data points were generated
using the RNS code \cite{Stergioulas:Morsink:1999, Stergioulas:Friedman:1995},
where a multipole extraction according to \cite{Pappas:Apostolatos:2012} was used.
The labels SLy, APR, FPS, and AU refer to equations of state considered in \cite{Chakrabarti:Delsate:Gurlebeck:Steinhoff:2013}.}
\label{S2plot}
\end{figure}

In Figure (\ref{S2plot}) the numerical value of $C_{ES^2}$ is shown as a function
of the spin length for fixed mass $\mu$ but different neutron star models.
It is apparent from the plot that
$C_{ES^2}$ is approximately independent of the spin length. However, one should
be careful and check this assumption for the specific case of interest.
This determination of $C_{ES^2}$ is actually a simple example of a matching
procedure. The quadrupole moment $J$ of the effective point particle is parametrized
through the ansatz (\ref{Mansatz}) as $J \sim C_{ES^2} S^2$. This is compared
(or matched) to the quadrupole moment of a numeric neutron star spacetime computed with
the RNS code \cite{Stergioulas:Morsink:1999, Stergioulas:Friedman:1995}.
Here the quadrupole moment is identified through the exterior spacetime. This
means that the effective point particle mimics the exterior spacetime
of a numerically constructed neutron star model, which depends crucially
on strong field effects in the interior. This makes $C_{ES^2}$ an interesting
indicator for both the neutron star equation of state and strong-field modifications
of gravity.
For black holes, a comparison with the Kerr metric leads to $C_{ES^2} = 1$.

Finally, we come back to the thermodynamic analogy to our approach.
The quadrupole relation $J \sim C_{ES^2} S^2$ can be viewed as a
simple (idealized) ``equation of state'' relating the macroscopic variables $J$
and $S$. As in the case of the ideal gas, this model can be improved to meet the
required accuracy. This can be done systematically here by extending the ansatz
(\ref{Mansatz}) to higher orders.

\subsection{Application\label{application}}
As an application for the spin-induced quadrupole constructed in the last section,
we consider the case of a test particle moving in a Kerr spacetime. This test
particle can be characterized as a pole-dipole-quadrupole particle. We aim
at an estimate for the relevance of the spin-squared contributions, so we may
consider a specific orbital configuration that simplifies the discussion.
This is obviously a circular orbit in the equatorial plane of the Kerr geometry.
Let us further assume that the spin of the test body is aligned with the rotation axis
of the background spacetime.

In the absence of a quadrupole, these orbits can be constructed in a simple manner,
which was first used in \cite{Rasband:1973}. This method is in fact still applicable
for the considered quadrupole model \cite{Steinhoff:Puetzfeld:2012}. It requires
that conserved quantities, spin supplementary condition,
and constraints on the orbital configuration are enough to uniquely fix the 10
dynamic variables contained in $p_{\mu}$ and $S^{\mu\nu}$. This is just an algebraic
calculation, in contrast to solving the differential equations of motion.
A numeric study for Schwarzschild spacetime is given in \cite{Bini:Geralico:2013}.

The spin supplementary condition ($S^{\mu\nu} p_{\nu} = 0$) contains three independent equations.
The constraint on the orbit provides three further independent conditions:
one due to equatorial orbits ($p^{\theta}=0$) and two due to spin alignment ($S^{\mu\theta} = 0$).
So we need to identify $10-3-3=4$ conserved quantities in order to solve for $p_{\mu}$ and $S^{\mu\nu}$
algebraically.
Three conserved quantities were already identified in Sec.\ \ref{conserved}.
These are the spin-length $S:= \sqrt{ \frac{1}{2} S_{ab} S^{ab} }$ and the quantities
derived from the two Killing vectors of Kerr spacetime ($\partial_t$ and $\partial_{\theta}$)
through (\ref{Killconst}). Well call the latter two the energy $E:=E_{\partial_t}$ and total angular
momentum $J_{\phi}:=E_{-\partial_{\phi}}$ of the particle.
The last remaining conserved quantity is just the mass-like parameter $\mu$, which in the
action approach is constant by assumption. However, one should remember that
(\ref{Mansatz}) is truncated and thus only approximately valid. One can equivalently
say that $\mu$ is only conserved approximately, corresponding to
the truncation of (\ref{Mansatz}). This point of view was taken in \cite{Steinhoff:Puetzfeld:2012}.

Now we are in a position to solve for $p_a$ and $S^{ab}$. Most important is the equation for $p^r$.
After some algebra \cite{Steinhoff:Puetzfeld:2012}, one finds that $(p^r)^2$ is given by a polynomial
of second order in $E$. We denote the roots of this polynomial by $U_{+}$ and $U_{-}$, i.e.,
\begin{equation}
\left(p^r\right)^2 \propto (E - U_{+}) (E - U_{-}) .
\end{equation}
For $p^r$ to be a real number, we need to have both $E \leq U_{+}$ and $E \leq U_{-}$,
or both $E \geq U_{+}$ and $E \geq U_{-}$. It turns out that the important relation is
just $E \geq U_{+}$ for the most relevant part of the parameter space. This justifies to
call $U_{+}$ effective potential: The test body can only move in the region where
$E \geq U_{+}$ and its turning points are given by $E = U_{+}$, because then
$p^r = 0$ (which implies $u^r = 0$, see \cite{Steinhoff:Puetzfeld:2012}). Therefore
the minimum of $U_{+}$ as a function of $r$ defines circular orbits. This completes
our construction.

\begin{figure}
\begin{center}
\includegraphics{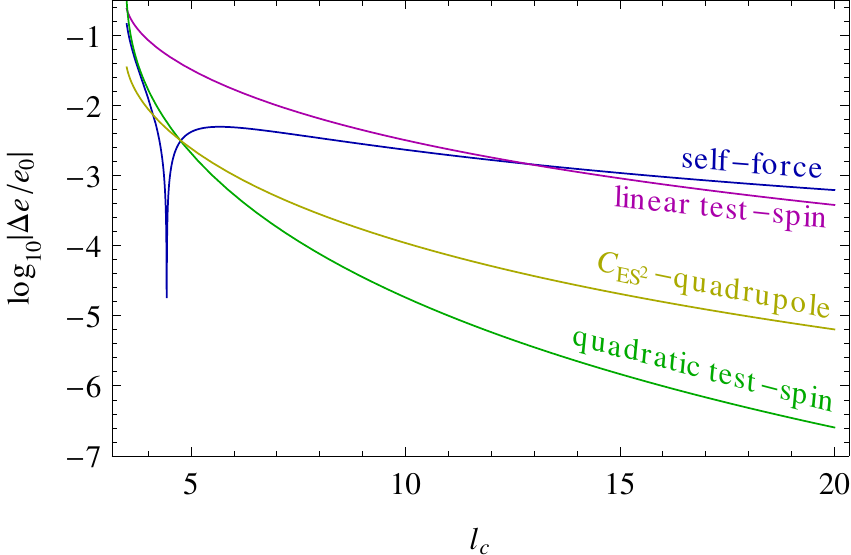}
\end{center}
\caption{Various corrections to the binding energy $e$ for a maximally spinning (small) black hole in a
Schwarzschild background. Here $l_c := J_{\phi} - S$ is the \emph{orbital} angular momentum.
The mass ratio is formally taken to be $q=1$ in the plot, though the result are only valid
for $q \ll 1$. The curves can be scaled to the case of interest ($q \lesssim 10^{-2}$):
self-force and linear spin effects scale as $\propto q$, the others as $\propto q^2$.}
\label{spinplot}
\end{figure}

The various contributions to the dimensionless binding energy $e := E/\mu - 1$
are plotted in Fig.\ \ref{spinplot}
for the case of a very rapidly rotating (small) black hole in a Schwarzschild background.
A comparison with recent results for the conservative part of the self-force
\cite{LeTiec:Barausse:Buonanno:2012} is also included.
In a Kerr background, the last stable circular orbit can be very close to the horizon,
so that the discussed effects can be some orders of magnitude stronger.
The reader is referred to \cite{Steinhoff:Puetzfeld:2012} for a more complete discussion.

\section{Dynamical quadrupole and tidal forces\label{Qdynamic}}
For the model developed in the last section, the quadrupole adiabatically
follows the spin evolution. Thus, the quadrupole is not an independent
dynamical variable. In this section, we are going to investigate dynamical
quadrupoles, but restrict to the nonspinning case for simplicity.

\subsection{Basic idea}
We have already discovered that the dynamical mass $\mathcal M$ plays a role similar
to a thermodynamic potential. From this perspective, one can compare the
variables it depends on, like $p_{\mu}$ and $S_{\mu\nu}$, to thermodynamic
state variables. Noticing that $p_{\mu}$ and $S_{\mu\nu}$ are the monopole
and dipole moment, a natural extension is to introduce dynamical
``state'' variables for other multipoles, too. A possible motivation arises
from the realization that stars have oscillation modes and that these modes
can be excited by tidal forces from an external time-dependent gravitational
field. This phenomenon is well understood in Newtonian gravity \cite{Press:Teukolsky:1977},
see also \cite{Alexander:1987, Rathore:Broderick:Blandford:2002, Flanagan:Racine:2007}
and references therein.
If one wants to capture it by our approach, one obviously must introduce
dynamical worldline variables corresponding to these oscillation modes.
Suitable point-particle actions were already discussed in \cite{Goldberger:Rothstein:2006:2,
Goldberger:Ross:2010}, though with applications to absorption or binary systems in mind.

The key to find a model for dynamical multipoles is to understand the
reaction of the multipoles to external fields.
We focus here on the response of the quadrupole to external tidal fields.
In fact, we will encode the quadrupole dynamics in terms of a response function.
This function can equivalently be called the propagator of the quadrupole
\cite{Goldberger:Rothstein:2006:2}, which better highlights the fact that
it is a necessary ingredient for deriving predictions using perturbative calculations, e.g.,
in the post-Newtonian approximation. A third possible naming is correlation
function between quadrupole and external field. This better accentuates
the parallels to statistical mechanics or thermodynamics. The idea is that
if one would be able to model the correlations of the most important multipoles
among each other and with external fields, then one can in principle predict
the motion of extended objects (with complicated internal structure) to any
desired precision.

It is important to notice that the multipole moments of a compact object can
be defined through their exterior field. The response functions of the multipoles
to externally applied tidal fields can therefore be obtained by analyzing the gravitational
field outside of the body. The final goal is to extract these functions from
numerical simulations of a \emph{single} compact object. However, for a first
simpler investigation one can restrict to linear perturbations
of nonrotating compact objects. The unperturbed metric in the exterior is then
just the Schwarzschild one. Because this metric is static and spherically symmetric, 
its linear perturbations can then be decomposed into Fourier basis in the time direction
and spherical harmonic basis $Y^{lm}(\theta, \phi)$ in angular directions. Then their radial dependence
is described by the famous Zerilli \cite{Zerilli:1970} or Regge-Wheeler \cite{Regge:Wheeler:1957}
equations for electric- or magnetic-parity-type perturbations, respectively.
The Zerilli equation can be transformed into the simpler Regge-Wheeler form
\cite{Chandrasekhar:1975}, so we can focus just on the latter one. It reads
\begin{equation}\label{ReggeWheeler}
\frac{d^2 X}{d r_*^2} + \left[ \left(1-\frac{R_S}{r} \right)
\frac{l (l+1) - \frac{3 R_S}{r}}{r^2} + \omega^2 \right] X = 0 ,
\end{equation}
where $\omega$ is the frequency of the perturbation, $l$ is the
angular momentum quantum number, $r$ is radial coordinate in the Regge-Wheeler gauge,
$R_S$ is the Schwarzschild radius (representing the mass of the body),
$r_* = r + R_S \log (r / R_S - 1)$ is the tortoise radial coordinate,
and $X$ denotes the Regge-Wheeler master function.
% and $S_X$ is the source (stress-energy) of the perturbation.
Given some boundary values for $X$
at the surface of the body (which result from a solution to the more complicated
interior perturbation equations), it is straightforward to integrate this
equation numerically. The question is how one can
decompose $X$ into external (applied) tidal field and multipolar field generated
by the body in response to the external field.
This is a complicated problem in the general relativistic case.
Let us therefore start with the Newtonian
theory in order to get a better understanding of the problem
\cite{Chakrabarti:Delsate:Steinhoff:2013:1}.

\subsection{Newtonian case}
The Newtonian case can be obtained as a weak field and slow motion
approximation of general relativity. That is, we have to set $R_S = 0$ (weak field)
and $\omega = 0$ (slow motion) in (\ref{ReggeWheeler}).
The perturbation of the Newtonian potential $\Phi_{\text{pert}}$ can be reconstructed as
\begin{equation}
\Phi_{\text{pert}} = - \frac{1}{2\pi} \int d\omega \sum_{lm}e^{i \omega t} Y^{lm}
        \frac{1}{2} \left[ \frac{d}{d r} + \frac{l (l+1)}{2 r} \right] X_{l m \omega},
\end{equation}
where the $X_{l m \omega}$ are solutions to the Newtonian limit of (\ref{ReggeWheeler})
for all values of the parameters $l$, $m$, and $\omega$.

The generic solution to the Newtonian limit of (\ref{ReggeWheeler}) reads
\begin{equation}\label{Nmatch}
X = C_1 r^{l+1} + C_2 r^{-l} ,
\end{equation}
where $C_1$ and $C_2$ are integration constants. The $r^{l+1}$ part diverges asymptotically,
which means that its source is located at infinity. Therefore, $C_1$ is the strength of the
external field. Similarly, the $r^{-l}$ part is singular at the origin and emanates from the
compact body, so $C_2$ describes the $l$-polar field of the body. The frequency-domain
response $\tilde{F}_l$ of the multipoles to external fields is then proportional to the ratio of
$C_2$ and $C_1$. In the conventions used in
\cite{Chakrabarti:Delsate:Steinhoff:2013:1, Chakrabarti:Delsate:Steinhoff:2013:2}, it holds
\begin{equation}\label{response}
\tilde{F}_l(\omega) = \frac{l (l-1)}{G (l+1) (l+2) (2l-1)!!} \frac{C_2}{C_1}
\end{equation}
This response must in general be computed numerically. The first step is to numerically
solve the interior problem of a perturbed body, including the interior gravitational field perturbation.
Then the gravitational field is matched to (\ref{Nmatch}) at the surface, which leads to
numeric values for the integration constants and thus for the response (\ref{response}).
This response can in general acquire a complicated frequency dependence through the
internal dynamics. Usually one defines normal oscillation modes by requiring that
the body keeps up a multipolar field without external excitation, i.e., for $C_1 = 0$.
Therefore the response (\ref{response}) has a pole at normal mode frequencies.

\begin{figure}
\begin{center}
\includegraphics{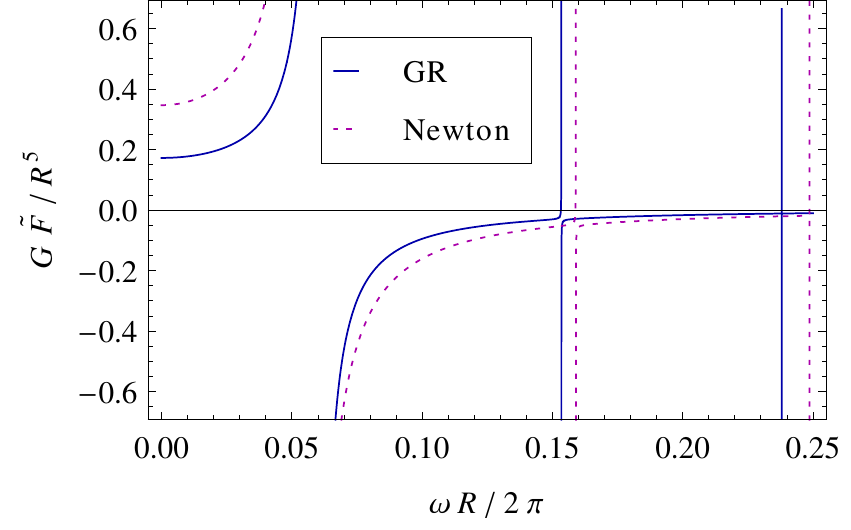}
\end{center}
\caption{Response function of the quadrupole, $l=2$, for a one solar
mass star. The equation of state is a polytrope with index 1 and
such that the radius $R$ is 17.7~km in the Newtonian case
or 15.7~km in the relativistic case.}
\label{propagatorplot}
\end{figure}

In the case of linear perturbations of a nonrotating barotropic star, the response
turns out to be quite simple. For the quadrupolar case $l=2$, the outcome is shown
in Fig.\ \ref{propagatorplot}. In fact, the form of the response can even be computed
analytically and reads \cite{Chakrabarti:Delsate:Steinhoff:2013:1}
\begin{equation}\label{Fanalytic}
\tilde{F}_l = \sum_n \frac{I_{nl}^2}{\omega_{nl}^2 - \omega^2} .
\end{equation}
This is just the sum of response functions of
harmonic oscillators with resonance frequencies (poles) at $\omega_{nl}$.
Here $n$ labels the type and overtone number of the oscillation modes. The constants $I_{nl}$
are the so called overlap integrals, which here simply take
the role of coupling constants between the oscillators and the external driving forces.
As a consequence, the internal dynamics can be captured by an effective action through
just a set of harmonic oscillators, which are coupled to the tidal force of the gravitational
field \cite{Chakrabarti:Delsate:Steinhoff:2013:1} (with coupling constants $I_{nl}$).
By fitting the numeric result for $\tilde{F}_l$ to (\ref{Fanalytic}), one can
extract the constants $\omega_{nl}$ and $I_{nl}$.

It is worth to point out that the presented Newtonian setup is simple enough
to perform explicitly the effective field theory procedure of integrating out
small scales, see \cite{Chakrabarti:Delsate:Steinhoff:2013:1}.
This turns a compact fluid configuration into a point particle on macroscopic
scales.

\subsection{Relativistic case at zero frequency}
Let us now return to the relativistic case, but restrict to even parity and the adiabatic case $\omega=0$.
The connection between the relativistic tidal constants defined in
\cite{Hinderer:2007, Damour:Nagar:2009:4, Binnington:Poisson:2009,
Bini:Damour:Faye:2012} and the response function is given by a Taylor-expansion,
\begin{equation}\label{Fexpand}
\frac{\tilde{F}_l(\omega)}{l!} = \mu_l + i \lambda_l \omega + \mu'_l \omega^2 + \Order(\omega^3) ,
\end{equation}
see \cite{Chakrabarti:Delsate:Steinhoff:2013:1}.
Here the constants $\mu_l$ are named after the astronomer A.\ E.\ H.\ Love, who
introduced them for tidal effects in the Earth-Moon system.
A dimensionless version of the Love numbers $\mu_l$ is often defined as
\begin{equation}
k_l = \frac{(2l-1)!!}{2 R^{2l+1}} G \mu_l ,
\end{equation}
where $R$ is the radius of the star.
The $\lambda_l$-term in (\ref{Fexpand}) is related to absorption \cite{Goldberger:Rothstein:2006:2}
and $\mu'_2$ was introduced in \cite{Bini:Damour:Faye:2012}.

\begin{figure}
\begin{center}
\includegraphics{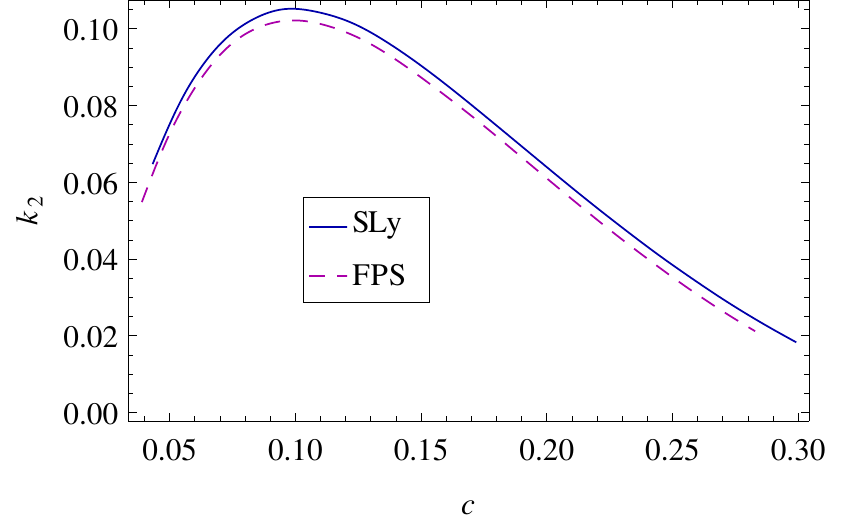}
\end{center}
\caption{Dimensionless quadrupolar Love number $k_2$ as a
function of the compactness $c = R_S / 2 R$ for two different equations of state (SLy, FPS).}
\label{k2plot}
\end{figure}

It remains to define how the response should be computed in the adiabatic
relativistic case. First, we again solve (\ref{ReggeWheeler}), this time
for $\omega=0$, and find an analytic result in terms of the Gauss hypergeometric
function ${}_2F_1$,
\begin{equation}
\begin{split}\label{Lovematch}
X &= C_1 r^{l+1} \; {}_2F_1(-l-2, 2 - l, -2 l ; R_S / r ) \nl
        + C_2 r^{-l} \; {}_2F_1(l-1, l+3, 2 (l+1) ; R_S / r ) ,
\end{split}
\end{equation}
see, e.g., \cite{Kol:Smolkin:2011}. Again we can obtain numeric values for
the integration constants by solving the perturbation equations inside the body
and then match the gravitational field to (\ref{Lovematch}) at the surface.
In the limit of $1 / r \rightarrow 0$
the hypergeometric functions are equal to $1$, so (\ref{Lovematch}) turns into
(\ref{Nmatch}). This implies that the interpretation of the integration
constants as magnitudes of external field and response is still valid.
The even-parity response in the adiabatic case $\omega=0$ then follows from
(\ref{response}) as before. A plot of the outcome in terms of the dimensionless
Love number $k_2$ is given in Fig.\ \ref{k2plot}. An extension of the application
from Sec.\ \ref{application} to adiabatic tidal deformations can be found in
\cite{Steinhoff:Puetzfeld:2012}.

For integer values of $l$, the hypergeometric functions in (\ref{Lovematch})
turn into polynomials (which possibly contain logarithms). Then one might worry
that the exponents on $r$ from the two independent solutions in (\ref{Lovematch})
can overlap and spoil an unique identification of external field and response.
However, this is avoided by examining $X$ for generic values of $l$, in the
sense of an analytic continuation.
This is in spirit similar to working in generic dimension, as done in
\cite{Kol:Smolkin:2011}.

% \begin{figure}
% \begin{center}
% \includegraphics{KerrPlotK2EOM13}
% \end{center}
% \caption{Corrections to the binding energy $e$ due to tidal deformation of a small extended body in a Kerr background. Curves must be scaled to the actual mass ratio, see text.}
% \label{Kerrplot}
% \end{figure}

\subsection{Relativistic case for generic frequency\label{Qdynamicfull}}
We now turn our attention to the case of generic frequency
in the even parity sector
\cite{Chakrabarti:Delsate:Steinhoff:2013:2}. One can still
solve (\ref{ReggeWheeler}) analytically \cite{Mano:Suzuki:Takasugi:1997},
this time in terms of a series involving hypergeometric functions.
We write the generic solution schematically as
\begin{equation}\label{MST}
X = A_1 X_{\text{MST}}^{l} + A_2 X_{\text{MST}}^{-l-1} ,
\end{equation}
where we denote the solution from \cite{Mano:Suzuki:Takasugi:1997}
by a subscript MST.

Note here that $X_{\text{MST}}^{l} \sim r^l$
and $X_{\text{MST}}^{-l-1} \sim r^{-l}$, which means that (\ref{response})
essentially still works. Of course, one has to take into account the normalization
of the $X_{\text{MST}}$ in order to rewrite the $C_i$ in (\ref{response})
in terms of the $A_i$.
This introduces complicated $\omega$-dependent corrections into (\ref{response}).
These are computed through a matching of the asymptotic field of the extended
body to the field of the point-particle model.
The details of this procedure can be found in \cite{Chakrabarti:Delsate:Steinhoff:2013:2}.
The basic steps are as follows:
\begin{itemize}
\item The field of the effective theory is obtained from an inhomogeneous version of
(\ref{ReggeWheeler}) with a point particle source. It is understood that
the post-Minkowskian approximation is applied, as this removes the singular point of
(\ref{ReggeWheeler}) at the Schwarzschild radius. The explicit form of the source term
derives from (\ref{EMT}).
\item The solution to the inhomogeneous equation is constructed from the
homogeneous solution (\ref{MST}) using the method of variation of parameters.
This method involves integrals over products of singular source and the $X_{\text{MST}}$.
The integration constants just represent a generic solution to the homogeneous
solution that can always be added.
\item Here the integration constants must be restricted further. Due to
the singular behavior of the differential equation at $r=0$, the homogeneous solution
might actually not be homogeneous at $r=0$. But the externally applied field is
homogeneous everywhere, including $r=0$.
The restriction of the integration constants is therefore equivalent
to the identification of the external part of the field and the part generated by the particle.
\item Notice that an $l$-pole source involves $l$ partial derivatives of a delta distribution.
This suggests to identify the self-field by $X_{\text{MST}}^{-l-1} \sim r^{-l}$ and
the external field by $X_{\text{MST}}^l \sim r^l$ for dimensional reasons.
Here the idea of analytic continuation in $l$ is again crucial.
\item The integrals arising in the variation of parameters are actually singular.
This is not surprising, as the self-field of point-particles always leads to this
kind of problem. A regularization method must be introduced.
\end{itemize}
These steps lead to a refined (frequency dependent) version of (\ref{response})
expressing the response function in terms of $A_1$ and $A_2$. The final step
is again to obtain numeric values for $A_1$ and $A_2$ for an actual (extended) neutron
star.

The result for the general relativistic response function is shown in
Fig.\ \ref{propagatorplot}. It can still be fitted by (\ref{Fanalytic})
very well. This implies that the internal dynamics can be approximated
by a set of harmonic oscillators. Restricting to the quadrupolar level
$l=2$ for simplicity, this translates to a dynamical mass of the form
\begin{equation}
\mathcal{M}^2 \approx \mu^2
+ \mu \sum_n ( \psi_n{}^{AB} \psi_{n AB}
        + \omega_n^2 \phi_{n AB} \phi_n{}^{AB}
        + I_n E^{AB} \phi_{n AB} ) ,
\end{equation}
where the internal dynamical variables $\phi_{n AB}$ and $\psi_n{}^{AB}$ only
have spatial components in the body-fixed frame ($\phi_{n 0 B} = 0 = \psi_n{}^{0B}$)
and are symmetric tracefree in the indices $A$ and $B$.
The dynamical equations for the quadrupolar worldline variables can
be extracted from (\ref{uLegendre}), (\ref{phidecompose}), and (\ref{LMwl}),
\begin{equation}
\dot{\phi}_n{}^{AB} =
        \frac{\alpha}{2} \frac{\partial \mathcal{M}^2}{\partial \psi_{n AB}} , \qquad
\dot{\psi}_{n AB} =
        - \frac{\alpha}{2} \frac{\partial \mathcal{M}^2}{\partial \phi_n{}^{AB}} .
\end{equation}
In the linear perturbation regime, the contributions of the internal dynamical
variables are small compared to $\mu^2$.
The index $n$ still labels the type of the oscillation mode. The mass quadrupole is the
coefficient in front of $E^{AB}$, i.e., $Q_{AB} := \sum_n I_n \phi_{n AB}$.
Now the frame enters through $E^{AB} = \lambda^A{}_{\mu} \lambda^B{}_{\nu} E^{\mu\nu}$,
so we need to check if (\ref{compat2}) is fulfilled. Using $\Lambda_0{}^{\mu} = r^{\mu}$
and $E_{\mu\nu} r^{\nu} = 0$ it is easy to see that this is the case.
In fact, (\ref{compat2}) is always fulfilled if the time direction of the body-fixed
frame $\Lambda_0{}^{\mu}$ drops out of the action.

% Recall that we intend to
% model an extended body by a multipolar point-particle. For point-particles
% one generally
% must utilize some regularization method in order to
% control divergent self-interactions.
Some final remarks on the problem of regularization of point particles are in order.
It was shown already in \cite{Goldberger:Ross:2010} that the quadrupole diverges
at order $\omega^2$ in dimensional regularization. It is therefore not surprising
that poles appear in the generalization of (\ref{response}) at order $\omega^2$,
which must be subtracted within some renormalization scheme. At the same time,
the poles give rise to an explicit appearance of a renormalization scale parameter,
which in a sense parametrizes the ambiguity in the choice of the renormalization
scheme. An important point is that this scale parameter is in fact fixed by the
requirement that the response function has an asymptotic behavior for
$\omega \rightarrow \infty$ compatible with (\ref{Fanalytic}). Different regularization
and renormalization schemes will in general lead to slightly different numeric values
for this scale parameter. However, within a given scheme its value can be uniquely
matched and is therefore not ambiguous.
In this sense, the regularization and renormalization scheme is a part of the
phenomenological model.

% \remark{Further explanations? redefinition of integrations constants is resummation.
% our choice is good due to the intuition from continuation in $l$.}

\section{Conclusions}
We considered point-particle models for extended bodies in gravity, in particular for
black holes and neutron stars. The multipoles of the point particles are adjusted
such that their field predicted from a weak field approximation matches
an exact/numerical solution for the extended object in question. This incorporates
strong field effects from the interior of the extended object in the model.
This is of particular importance when binary systems are considered using
weak field approximations, e.g., for gravitational wave source modeling or
pulsar timing.

Therefore, point-particle actions are far more powerful than what was probably envisioned
when they were first investigated \cite{Westpfahl:1969:2, Bailey:Israel:1975}.
The resulting equations of motion are similar to Dixon's results.
Here we developed astrophysical realistic models for the multipoles in these equations.
The latest development is the inclusion of oscillation modes in relativistic tidal interaction
of neutron stars.

An interesting topic not discussed here are universal relations for various neutron
star properties. Here ``universal'' refers to an approximate independence among various
proposed realistic equations of state. In \cite{Yagi:Yunes:2013:1, Yagi:Yunes:2013:2}
universal relations between the dimensionless moment of inertia
$I / G^2 \mu^3$, the quadrupolar Love number $\mu_2 / G^4 \mu^5$, and the
quadrupole constant $C_{ES^2}$ were found and coined I-Love-Q relations.
Further investigations, also including
higher multipoles, followed shortly afterwards
\cite{Maselli:Cardoso:Ferrari:Gualtieri:Pani:2013, Baubock:Berti:Psaltis:Ozel:2013,
Haskell:etal:2013, Doneva:etal:2013, Pappas:Apostolatos:2013, Yagi:2013}.
This indicates that coefficients in (\ref{Mansatz}) arising at higher orders
are actually not independent, but are (approximately) fixed by universality.
(For black holes, in fact all coefficients are fixed, which is guaranteed by
the no hair theorem.) This makes the expansion (\ref{Mansatz}) a meaningful
tool to study the impact of the equation of state on observations, as predictions
of the effective model are then parametrized by only a small set of constants.

The most interesting development for the future is probably the
description of oscillation modes for rotating bodies, which can be tried in
a slow rotation approximation. It is also interesting to investigate if universal properties
hold for the ingredients of the response function, e.g., for the overlap integrals.

\paragraph{Note added in arXiv version:}
An action for a dynamical quadrupole in the Newtonian gravity is also given in
Ref.~\cite{Flanagan:Hinderer:2007} which was missed.
In Ref.~\cite{Marsat:2014} the equations of motion for the center of mass were
obtained in a manifestly covariant manner using a family of wordlines and
explicit expressions for the equations of motion including all gravitational
multipoles are given in the appendix. A treatment of the spin supplementary
condition improving on Sec.~\ref{SSCcond} here is given in Refs.~\cite{Levi:Steinhoff:2015:2, Steinhoff:2015}.

\paragraph{Acknowledgements}
I am indebted to all of my collaborators contributing directly or indirectly to the
material presented here:
Sayan Chakrabarti, T{\'e}rence Delsate, Norman G{\"u}rlebeck, Johannes Hartung, Steven Hergt,
Dirk Puetzfeld, Gerhard Sch{\"a}fer, and Manuel Tessmer.
This work was supported by DFG (Germany) through projects STE 2017/1-1 and STE 2017/2-1,
and by FCT (Portugal) through projects
SFRH/BI/52132/2013 and PCOFUND-GA-2009-246542 (co-funded by Marie Curie Actions).

%\bibliographystyle{unsrt}
% COMMENTED: alternative bibstyle
%\bibliographystyle{utphys}
%\bibliography{steinhoff_eom_proceedings_2013}

\providecommand{\href}[2]{#2}\begingroup\raggedright\endgroup

\end{document}